\documentclass[pdftex,5p,twocolumn]{elsarticle}%[review,1p]{elsarticle}
\usepackage{txfonts}
\usepackage{natbib}
\bibpunct{(}{)}{;}{a}{}{,}
\usepackage{indentfirst} % to indent the first line of a section
\usepackage{amssymb}
\usepackage{array}
\usepackage{verbatim}

\journal{Icarus}

\begin{document}

\begin{frontmatter}

%%%%%%%%%%%%%%%%%%%%%%%%%%%%%%%%%%%%%%%%%%%%%%%%

\title{Water transport in protoplanetary disks and the hydrogen isotopic composition of chondrites}

\author[cita,lmcm]{Emmanuel Jacquet\corref{cor1}}
\ead{ejacquet@cita.utoronto.ca}
\author[lmcm]{Fran\c{c}ois Robert}
%\author{Emmanuel Jacquet\inst{1} \and Matthieu Gounelle\inst{1} \and S\'{e}bastien Fromang\inst{2,3}}

\cortext[cor1]{Corresponding author}

\address[cita]{Canadian Institute for Theoretical Astrophysics, University of Toronto, 60 St Georges Street, Toronto, ON M5S 3H8, Canada}
\address[lmcm]{Laboratoire de Min\'{e}ralogie et de Cosmochimie du Mus\'{e}um, CNRS \& Mus\'{e}um National d'Histoire Naturelle, UMR 7202, 57 rue Cuvier, 75005 Paris, France.}

\begin{abstract}

The D/H ratios of carbonaceous chondrites, believed to reflect the hydrogen isotopic composition of water in the inner early solar system, are intermediate between the protosolar value and that of most comets. The isotopic composition of cometary water has been accounted for by several models where the isotopic composition of water vapor evolved by isotopic exchange with hydrogen gas in the protoplanetary disk. However, the position and the large range of variation of the distribution of D/H ratios in carbonaceous chondrites have yet to be explained. In this paper, we assume that the D/H composition of cometary ice was achieved in the disk building phase and model the further isotopic evolution of water in the inner disk in the classical T Tauri stage. Reaction kinetics compel isotopic exchange between water and hydrogen gas to stop at $\sim$500 K, well inside the snow line. However, the equilibrated water can be transported to the snow line (and beyond) via turbulent diffusion and consequently mix with isotopically comet-like water. Thus the competition between outward diffusion and net inward advection established an isotopic gradient, which is at the origin of the large isotopic variations in the carbonaceous chondrites and other water-bearing objects accreted in the protoplanetary disk.

  Under certain simplifying assumptions, we calculate analytically the probability distribution function of the D/H ratio of ice accreted in planetesimals and compare it with observational data. The distribution is found to essentially depend on two parameters: the radial Schmidt number Sc$_R$, which ratios the efficiencies of angular momentum transport and turbulent diffusion, and the range of heliocentric distances over which currently sampled chondrite parent bodies were accreted. The minimum D/H ratio of the distribution corresponds to the composition of water condensed at the snow line, which is a function of both the composition of equilibrated water having diffused from hotter disk regions and the efficiency of this outward transport as measured by Sc$_R$. Observations constrain the latter to low values (0.1-0.3), which suggests that turbulence in the planet-forming region was hydrodynamical in nature, as would be expected in a dead zone. Such efficient outward diffusion would also account for the presence of high-temperature minerals in comets.
 \end{abstract}


\end{frontmatter}

\section{Introduction}

  The isotopic composition of hydrogen (as expressed by the D/H ratio) is a valuable tracer of the origin of water in the solar system \citep{Robert2006}. Primitive meteorites (\textit{chondrites}), presumably the building blocks of terrestrial planets, allow a glimpse at the D/H ratio of water in the protoplanetary disk that surrounded our Sun 4.57 Ga ago. Indeed, many chondrites, in particular \textit{carbonaceous chondrites}, contain clays formed through alteration of originally anhydrous silicates by water presumably incorporated as ice along with rock during accretion \citep{Brearley2003,Ghoshetal2006}. 

 The measured D/H ratios of bulk carbonaceous chondrites\footnote{And carbonaceous chondrite-like microclasts in howardites \citep{Gounelleetal2005}.} (see Fig. \ref{histogram}), generally thought to reflect that of accreted water (but see \citealt{Alexanderetal2012}), span a range of $120\times 10^{-6}$ to $230\times 10^{-6}$ (excluding CR chondrites). The distribution, which is skewed to heavy isotopic compositions, has a mean $(156\pm 3)\times 10^{-6}$ close to the $(149\pm 3)\times 10^{-6}$ estimated for the bulk Earth \citep{Lecuyeretal1998} % (SMOW $155.8\times 10^{-6}$)
 --- consistent with a chondritic source for terrestrial water. The D/H ratios of carbonaceous chondrites is systematically lower than that exhibited by many Oort-cloud comets ($(296\pm 25)\times 10^{-6}$; \citealt{Hartoghetal2011}), although D/H ratios of $(161\pm 24)\times 10^{-6}$ and $(206\pm 22)\times 10^{-6}$ have been determined for Jupiter-family comet Hartley 2 and Oort-cloud comet Garradd, respectively \citep{Hartoghetal2011, BockeleeMorvanetal2012}. The composition of interplanetary dust particles, though broadly similar to that of carbonaceous chondrites \citep[e.g.][]{EngrandMaurette1998, Bradley2005}, has a significant tail extending to cometary values and beyond. Both chondritic and cometary domains of variation are markedly distinct from both the estimated protosolar value ($(20\pm 3.5)\times 10^{-6}$; \citealt{GeissGloeckler2003}), dominated by the isotopic composition of hydrogen gas, and the high values (D/H $\gtrsim 10^{-3}$) determined in molecular clouds for molecules other than H$_2$, consistent with predictions from ion-molecule reactions \citep[see e.g.][]{Robert2006}. However, high D/H values $\gtrsim 2\times 10^{-3}$ %0.0036
have been reported at the micrometer scale in clays of the Semarkona chondrite \citep{Pianietal2012, Piani2012} and may reflect the composition of pristine interstellar ice grains.

\begin{figure}
\resizebox{\hsize}{!}{
\includegraphics{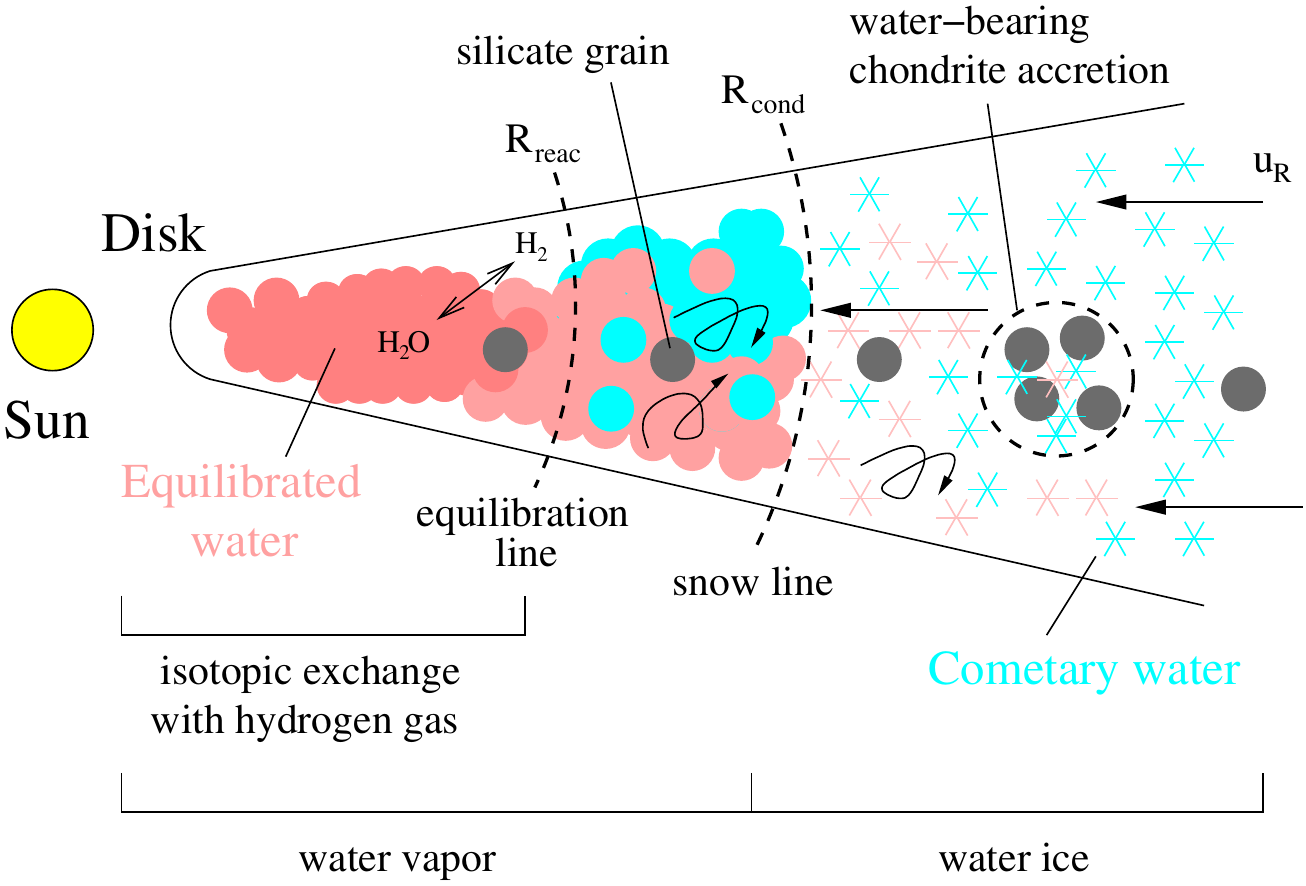}
}
\caption{Cartoon of the model investigated here. We distinguish between (comet-like) ``cometary water'' (cyan) and ``equilibrated water'' (pink) that has undergone isotopic exchange with the hydrogen gas inside the ``equilibration line'' ($R_{\rm reac}$). Water is gaseous inside the snow line ($R_{\rm cond}$) and solid beyond (as symbolized by the cloud and snowflakes symbols, respectively). Accretion of water as ice occurs beyond the snow line. Arrows symbolize motions of the gas.}
\label{cartoon}
\end{figure}

  In the protoplanetary disk, the D/H ratio of water evolves mainly through isotopic exchange with the hydrogen gas (whose composition remains essentially fixed at the protosolar value because it contains the bulk of the H of the system), that is, via the reaction: 

\begin{equation}
\rm HDO + H_2 \leftrightarrows H_2O + HD
\label{reaction}
\end{equation}

The equilibrium fractionation factor (D/H)$_{\rm H_2O}$/(D/H)$_{\rm H_2}$ is unity at high temperatures ($\gtrsim 1000$ K), but increases with decreasing temperature \citep{Richetetal1977}.

 Using one-dimensional disk models, \citet{Drouartetal1999} showed that reaction (\ref{reaction}) was unable to account for the chondritic or cometary D/H ratios if water was assumed to have formed in the protoplanetary disk with the protosolar D/H value. Indeed reaction kinetics would have been prohibitively slow in those cold regions of the disk where sufficient D enrichment would have been predicted by thermodynamic equilibrium. On the other hand, \citet{Drouartetal1999} found that cometary compositions could be obtained if water was inherited from the protosolar cloud (with heavy D/H $\sim 10^{-3}$). Water would have evolved toward isotopically lighter compositions in the inner disk and diffused in the colder regions owing to turbulence, a result also obtained by \citet{Mousisetal2000} and \citet{Hersantetal2001}. Transport and mixing would have been particularly efficient in the early stages of disk evolution, if the disk was first built compact and then expanded because of turbulence \citep{Yangetal2012}. Such a picture is also consistent with the presence of crystalline silicates in comets \citep{BockeleeMorvanetal2002}.

  While these previous works focused on reproducing D/H value observed in comets, implications on the composition of water in the inner solar system, and in particular the fairly large domain of variations of the D/H ratio in chondrites have yet to be worked out in this framework. This is the purpose of this paper.

  In this paper, we consider a disk model pertaining to the classical T Tauri phase, that is, after infall has ceased and the disk radius has become large compared to the heliocentric distances of interest. This is the epoch where chondrite accretion is believed to have taken place. At that time, the D/H ratio of water ice in the outer disk is assumed to have been already set and homogenized at the value of $\sim 300\times 10^{-6}$ measured in most comets as a result of early disk processes (this would correspond to the plateau in the simulation of \citet{Yangetal2012}, whose level varies little after infall has stopped). Close to the Sun, isotopic exchange between (gaseous) water and hydrogen gas (reaction (\ref{reaction})) is still going on. However, beyond a certain heliocentric distance (corresponding to a temperature of $\sim$500 K, as we shall argue later), the kinetics of this reaction are so slow that essentially no isotopic exchange occurs there. Due to gas turbulence, however, some of the water equilibrated with H$_2$ sunward of this heliocentric distance will be transported outward and will reach the \textit{snow line}, where ice condenses. Water beyond the snow line---sampled by chondrites---will thus be a mixture of two components: equilibrated water having diffused from the hot inner regions of the disk, and cometary water drifting inward from the outer disk.

  A radial gradient in D/H ratio is thus established by the competition between outward diffusion and inward advection which sets the proportions of equilibrated water and cometary water as a function of heliocentric distance. The range of D/H ratios of water-bearing bodies thus extends from cometary compositions down to the composition of water condensed at the snow line. The composition sampled at the snow line is hence intermediate between cometary values and those expected from equilibration with H$_2$. This schematic description provides a framework to account for the fairly large variations of the D/H ratio in water of solar system bodies. % As accretion is more efficient at the snow line than at larger heliocentric distances, the D/H of chondritic water, assumed to have been incorporated as ice, will be typically close to that of water condensed at the snow line. This value is dictated by both reaction kinetics in the hot regions of the disk and transport efficiency from there to the snow line. 
A cartoon of the scenario is presented in Fig. \ref{cartoon}.

  The goal of this paper is to calculate analytically the distribution of D/H ratios of chondritic water that results from this picture and compare it to the observational data, in order to constrain the model parameters. In particular, the calculation will be constrained by the asymmetrical shape of the distribution and its position relative to cometary values. We stress that here the cometary value is a \textit{starting}, observationally given parameter of the model. Our model does not aim at reproducing the composition of comets in contrast to the studies of \citet{Drouartetal1999, Mousisetal2000, Hersantetal2001, Yangetal2012} ; rather, in complementarity to those, it focuses on chondrite parent bodies, that is, the inner regions of the solar system. Also in complementarity to these numerical studies, our work, in being analytic in nature, allows us to nail down, under certain simplifying assumptions, the relevant parameters (essentially two) that govern the distribution of chondritic D/H ratios, namely: the range of heliocentric distances sampled by chondrites and the radial Schmidt number---which essentially ratios the efficiencies of angular momentum transport and turbulent diffusion. We will find that low values of the radial Schmidt number yield D/H variations consistent with observations. The paper is organized as follows: We describe the model assumptions in Section \ref{Modeling}, present and discuss the results in Sections \ref{Results} and \ref{Discussion}, respectively. In Section \ref{Conclusion}, we conclude. For the sake of clarity, specific derivations are deferred to appendices.

\section{Modeling}
\label{Modeling}

In this section, we introduce our notations and modeling principles. We successively consider the disk model (Section \ref{The disk}), the transport of water (Section \ref{Water transport}), its D/H ratio (Section \ref{D/H ratio of water}) and our prescription for accretion and delivery to Earth (Section \ref{Accretion and delivery to Earth of meteoritic water}).

\subsection{The disk}
\label{The disk}
We consider an axisymmetric, vertically isothermal turbulent disk. The disk is assumed to have a stationary surface density profile in its inner regions such that its (radially constant) mass accretion rate $\dot{M}$ is given by:
\begin{equation}
\dot{M}\equiv-2\pi R\Sigma u_R = 3\pi\Sigma\alpha\frac{c_s^2}{\Omega}
\label{steady mass accretion rate}
\end{equation}
with $R$ the heliocentric distance, $\Sigma$ the surface density, $u_R$ the vertically averaged radial velocity of the gas, $\alpha$ the vertically averaged turbulence parameter, $\Omega$ the Keplerian angular velocity and $c_s=\sqrt{k_BT/m}$ the isothermal sound speed---where $k_B$ and $m$ are the Boltzmann constant and the mean molecular mass% (here 2.33 times the proton mass)
, respectively, and $T$ the temperature. The steady-state approximation is expected to hold so long the viscous evolution timescale,
\begin{equation}
t_{\rm vis}(R)\equiv\frac{R^2}{3\alpha c_s^2/\Omega}=0.05\:\mathrm{Ma}\:R_{\rm AU}^{1/2}\left(\frac{300\:\mathrm{K}}{T}\right)\left(\frac{10^{-3}}{\alpha}\right),
\label{tvis}
\end{equation}
is shorter than the disk's age in the regions of interest. Similarly to \citet{Drouartetal1999} and \citet{Hersantetal2001}---see also \citet{Hartmannetal1998} and \citet{Chambers2009}---, we take the mass accretion rate to evolve in time as\footnote{This is the evolution expected from the self-similar solution of the continuity equation for $\alpha c_s^2/\Omega\propto R$ (see e.g. \citealt{Garaud2007}) as is the case here in the outer disk.}
\begin{equation}
\dot{M}=\frac{\dot{M}_0}{\left(1+t/t_0\right)^{3/2}}
\end{equation}
where $\dot{M}_0$ is the mass accretion rate at some initial time $t=0$ %(taken to be the start of chondrite accretion)
 and $t_0$ an evolution timescale ($t_0>t_{\rm vis}(R)$). %(which we will not actually need to specify).

  We adopt the same temperature prescription as in \citet{Jacquetetal2012S}:
\begin{equation}
T=\mathrm{max}\bigg[\left(\frac{3}{128\pi^2}\frac{\kappa m}{\sigma_{\rm SB}k_B\alpha}\dot{M}^2\Omega^3\right)^{1/5}, f_T T_0R_{\rm AU}^{-1/2}  \bigg],
\label{T steady}
\end{equation}
with $\kappa$ the specific opacity, $\sigma_{\rm SB}$ the Stefan-Boltzmann constant, $T_0=280$ K, $f_T$ a dimensionless constant parameter (for which we will adopt a fiducial value of 0.5) and $R_{\rm AU}\equiv R/(1\:\rm AU)$. This prescription essentially states that inner disk regions are dominated by the dissipation of turbulence (``viscous heating''), which decreases with decreasing accretion rate, whereas the outer disk regions are dominated by reprocessing of solar radiation. Hence, the snow line (at heliocentric distance $R_{\rm cond}$), which corresponds to a fixed temperature of $T_{\rm cond}$=170 K, recedes toward the Sun with the passage of time.

 For simplicity, we will assume that $\alpha$ and $\kappa$ are constant throughout the disk's extent and evolution.

\subsection{Water transport}
\label{Water transport}

We now turn to the transport of water. We distinguish between \textit{nebular} and \textit{accreted} water. \textit{Nebular water} is water dynamically coupled to the gas, whether as water vapor or ice-bearing grains%\footnote{Possibly also as mantles around silicate dust particles.}
, and behaves as part of the disk. It is predominantly gaseous inside the snow line and solid outside it. \textit{Accreted water} refers to water retrieved from the gas by incorporation in %meter-sized boulders, 
planetesimals or comets.

 We first focus on the dynamics of nebular water. Since it behaves as a passive contaminant in the gas, the evolution equation of its surface density $\Sigma_{\rm H_2 O}$ reads \citep[e.g.][]{CieslaCuzzi2006}:
\begin{eqnarray}
\frac{\partial\Sigma_{\rm H_2O}}{\partial t}+\frac{1}{R}\frac{\partial}{\partial R}\left[R\left(\Sigma_{\rm H_2 O}u_R-\delta_R\frac{c_s^2}{\Omega}\Sigma\frac{\partial}{\partial R}\left(\frac{\Sigma_{\rm H_2 O}}{\Sigma}\right)\right)\right]\nonumber\\
=-S_{\rm coll}%S_{\rm shatter}-S_{\rm coag}
\label{water evolution}
\end{eqnarray}
with $S_{\rm coll}$ the sink term resulting from collisions --- both coagulation and shattering processes (contributing positively and negatively to it, respectively) ---, and $\delta_R$ a dimensionless number (of order $\alpha$) parameterizing turbulent diffusion. We neglect any gas-solid drift induced by gas drag (we shall return to this assumption in Section \ref{Caveats}).

 If coagulation and shattering can be ignored, it is straightforward to see that $\Sigma_{\rm H_2 O}\propto \Sigma$ is a solution of the equation% (with a proportionality constant dictated by cosmic abundances)
. This should remain a reasonable approximation for coagulation timescales longer than $t_{\rm vis}$. There is evidence suggesting that accretion was indeed quite inefficient/slow in the early solar system: age dating of chondrite components and studies of the thermal evolution of their parent bodies are consistent with accretion being a protracted process on a few-Ma timescale \citep[e.g.][]{Villeneuveetal2009,GrimmMcSween1989,Kleineetal2008, Connellyetal2012} and the total mass of the current planetary system (see e.g. \citealt{Hayashi1981}) is one order of magnitude lower than the nonvolatile content of disks at the end of the infall phase \citep[e.g.][]{YangCiesla2012} assuming a solar metallicity. We shall thus assume that accretion was inefficient and ignore its feedback on the dynamics of nebular water so that $\epsilon_{\rm H_2 O}\equiv\Sigma_{\rm H_2 O}/\Sigma$ can be considered constant (we shall return to this issue in Section \ref{Discussion}).

\subsection{Hydrogen isotopic composition of water}
\label{D/H ratio of water}

  In this paper, we assume that the D/H ratio of water is dictated by isotopic exchange between originally isotopically heavy (comet-like) water and isotopically light hydrogen gas \citep[e.g.][]{Drouartetal1999, Mousisetal2000, Hersantetal2001}. To make the problem analytically tractable, we circumvent an accurate treatment of reaction kinetics by schematically distinguishing between two types of water: \textit{cometary water} is water that has never experienced temperatures in excess of a ``reaction temperature'' $T_{\rm reac}$ during the classical T Tauri stage of the disk, and is assumed to retain a relatively heavy D/H ratio denoted (D/H)$_h$%. In detail, this ``cometary water'' may result from the mixing of isotopically diverse contributions (if never vaporized; L. Piani, private communication), as suggested by the range of D/H ratios encountered at small scales in unequilibrated ordinary chondrites and CR chondrites \citep{DelouleRobert1995} but our model only depends on the average value $x_h$. 
; \textit{equilibrated water}, on the other hand, is water that did experience temperatures above $T_{\rm reac}$ and therefore has a D/H ratio set to a fixed value (D/H)$_l$ through isotopic exchange with hydrogen gas (a justification for this treatment based on reaction kinetics is provided in \ref{kinetics}). $T_{\rm reac}$, (D/H)$_l$ and (D/H)$_h$ are fixed parameters of the model and are assigned values of 500 K, $40\times 10^{-6}$ and $300\times 10^{-6}$ for them, respectively (see \ref{kinetics}). The value of $40\times 10^{-6}$ corresponds to the maximum D/H ratio that can be reached by water in equilibrium with molecular hydrogen. Indeed, although enrichment in D could be in principle higher at lower equilibration temperature, slow kinetics of the isotopic exchange reaction prevent this equilibrium from being actually attained below $T_{\rm reac}$.

  We denote by $R_{\rm reac}$ the heliocentric distance where $T=T_{\rm reac}$, which is given by:
\begin{eqnarray}
R_{\rm reac} & = &\left(\frac{3\kappa m \dot{M}^2(GM_{\odot})^{3/2}}{128\pi^2\sigma_{\rm SB}k_B\alpha T_{\rm reac}^5}\right)^{2/9}\nonumber\\
& = & 1\:\mathrm{AU}\dot{M}_{-8}^{4/9}\left(\frac{\kappa}{0.5\:\rm m^2/kg}\frac{10^{-3}}{\alpha}\right)^{2/9}\left(\frac{500\:\rm K}{T_{\rm reac}}\right)^{10/9},
%\mathrm{max}\left( \left(\frac{3\kappa m \dot{M}^2\Omega_0^3}{128\pi^2\sigma_{\rm SB}k_B\alpha T_{\rm reac}^5}\right)^{2/9}%  ,\left(\frac{f_T T_0}{T_{\rm reac}}\right)^2\right)
\label{Rreac}
\end{eqnarray}
with $\dot{M}_{-8}\equiv \dot{M}/(10^{-8}\:\rm M_{\odot}\cdot a^{-1})$, in the viscous-dominated regime. The mass accretion rate where $R_{\rm reac}$ enters the irradiation-dominated regime (see equation (\ref{Rbr})) is in our model:
\begin{eqnarray}
\dot{M}_{\rm reac} & = & \left(\frac{128\pi^2\sigma_{\rm SB}k_B\alpha(f_TT_0)^9}{3\kappa m \Omega_0^3T_{\rm reac}^4}\right)^{1/2}\nonumber\\
& = & 4\times 10^{-11}\:\mathrm{M_{\odot}.a^{-1}}\left(\frac{\alpha}{10^{-3}}\frac{0.5\:\mathrm{m^2/kg}}{\kappa}\right)^{1/2}\left(\frac{f_T}{0.5}\right)^{9/2}\nonumber\\
& & \left(\frac{500\:\mathrm{K}}{T_{\rm reac}}\right)^2
\end{eqnarray}
which is a very low value (compared e.g. to mass accretion rates reported by \citealt{Hartmannetal1998}) so the disk is likely to have largely dissipated by that time. Moreover, it may be seen from equation (\ref{Rbr}) that then $R_{\rm reac}\lesssim 0.1$AU, most likely inside the inner edge of the disk. So certainly at that time the validity of our disk model has broken down. We will find it convenient to take the corresponding time $t_{\rm reac}$ as the final time in our calculation.

  Outside $R_{\rm reac}$, the surface density of equilibrated water $\Sigma_{\rm H_2 O, eq}$ is governed by the same equation as total water, i.e. equation (\ref{water evolution}) with $\Sigma_{\rm H_2 O}$ replaced by $\Sigma_{\rm H_2 O, eq}$. A stationary solution to this equation, expected to be attained by $t_{\rm vis}$, is given by \citep[][ see also \ref{appendX}]{ClarkePringle1988, Stevenson1990}
\begin{equation}
\frac{\Sigma_{\rm H_2 O, eq}}{\Sigma_{\rm H_2 O}}=\left(\frac{R_{\rm reac}}{R}\right)^{3\textrm{\scriptsize Sc}_R/2}
\label{X eau}
\end{equation}
where we have introduced the radial Schmidt number
\begin{equation}
\textrm{Sc}_R=\frac{\alpha}{\delta_R}.
\end{equation}
We will henceforth adopt this solution. In other words, we assume that the radial distribution of equilibrated water undergoes a quasi-static evolution as the mass accretion rate (and thus $R_{\rm reac}$) decreases.

\begin{figure}
\resizebox{\hsize}{!}{
\includegraphics{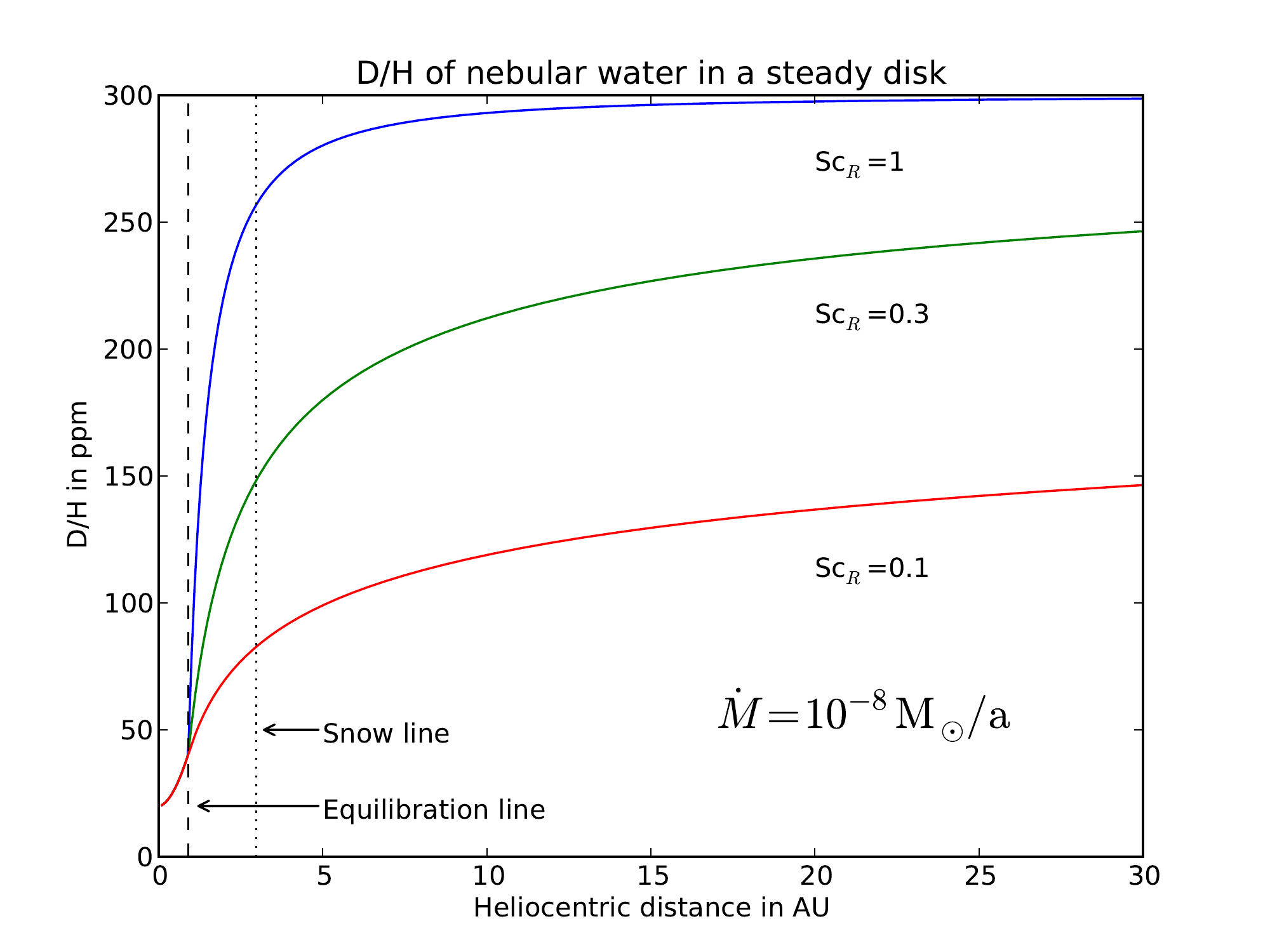}
}
\caption{D/H ratio (expressed in parts per million (ppm)) of water as a function of heliocentric distance in a steady disk, for three values of the radial Schmidt number Sc$_R$. In the ``reaction zone'', delimited by a vertical dashed line (``equilibration line''), equilibrium isotopic fractionation with hydrogen gas is assumed ; beyond, mixing with cometary water controls the D/H ratio (equation (\ref{x(R)})). We have taken $\dot{M}=10^{-8}\:\mathrm{M_\odot/a}$, $\alpha=10^{-3}$, $\kappa=0.5\:\mathrm{m^2/kg}$, $f_T=0.5$. The smaller the Sc$_R$, the more efficient radial mixing is. The position of the snow line (where water condenses) is indicated by a vertical dotted line: water is gaseous inside the snow line and solid outside.}
\label{D/H vs R}
\end{figure}

\begin{figure}
\resizebox{\hsize}{!}{
\includegraphics{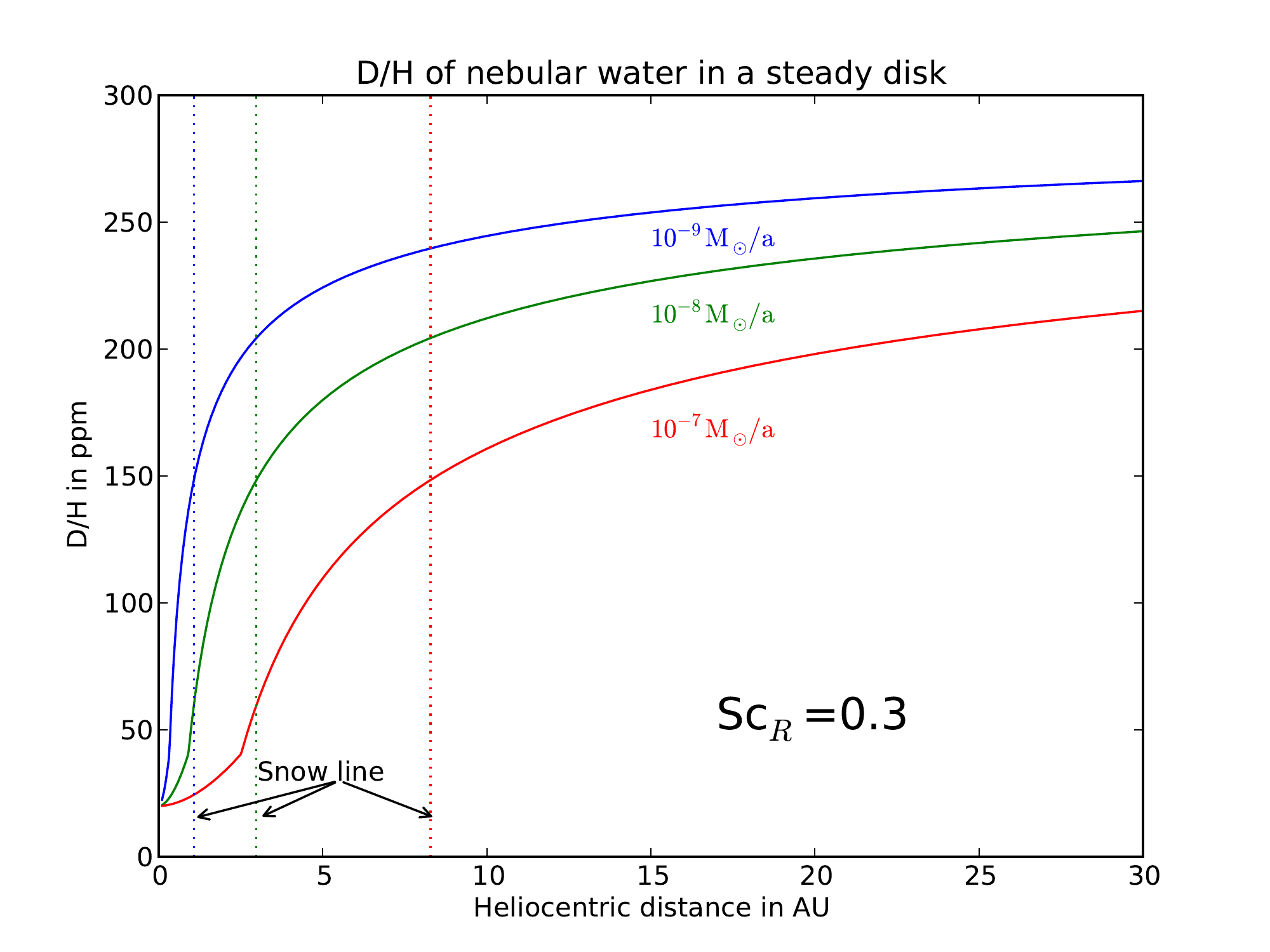}
}
\caption{Same as Fig. \ref{D/H vs R} but for a fixed Sc$_R$=0.3 and with varying mass accretion rates.
}
\label{D/H vs R Mdot}
\end{figure}

  In this case, the D/H ratio of the total nebular water is given by\footnote{We use the fact that D/H$\ll 1$.}:
\begin{equation}
\left(\frac{\mathrm{D}}{\mathrm{H}}\right)(R,t)=\left(\frac{\mathrm{D}}{\mathrm{H}}\right)_h-\left(\left(\frac{\mathrm{D}}{\mathrm{H}}\right)_h-\left(\frac{\mathrm{D}}{\mathrm{H}}\right)_l\right)\left(\frac{R_{\rm reac}(t)}{R}\right)^{3\textrm{\scriptsize Sc}_R/2}.
\label{x(R)}
\end{equation}
Thus, D/H is a monotonically increasing function both of heliocentric distance and time (as over time $\dot{M}$ and hence $R_{\rm reac}$ decrease). This is plotted in Fig. \ref{D/H vs R} and \ref{D/H vs R Mdot}. %For any given time $t$, this equation may be inverted to express $R(x,t)$ as:
%\begin{equation}
%R(x,t)=R_{\rm reac}(t)\left(\frac{x_h-x_l}{x_h-x}\right)^{2/3\textrm{Sc}_R}
%\end{equation}
%For a fixed $x$, this is a decreasing function of time.

%Note : boundary conditions of Hersant & Cie is dx/dR=0 at min(RD,Rcond)!

\subsection{Accretion and delivery to Earth of water-bearing chondrites}
\label{Accretion and delivery to Earth of meteoritic water}

At this point, we have wholly prescribed the isotopic and transport properties of nebular water in our model. We have yet to relate this nebular water to the D/H distribution of chondrites.
 
% SI, epsilon*Omega

  We first need to make a prescription for coagulation/shattering, which determines the rate at which water is incorporated in chondritic bodies. Similarly to \citet{Cassen1996} (see also \citealt{CieslaCuzzi2006, Weidenschilling2004}), we posit a collision term of the form
\begin{equation}
S_{\rm coll}(R,t)=\frac{\Sigma_{\rm H_2 O}}{t_{\rm coag}(R)}\theta(R-R_{\rm cond}(t)),
\end{equation}
with the coagulation timescale $t_{\rm coag}(R)$ taken to scale like the local orbital timescale ($\Omega^{-1}$), $\theta$ the Heaviside function defined by
\begin{equation}
\theta(z)=
\left\{\begin{array}{rr}
0\:\mathrm{if}\:z<0\\
1\:\mathrm{if}\:z\geq 0
\end{array}\right.
\end{equation}
We thus consider that water is accreted solely as ice. 

  We finally need to prescribe the delivery of accreted material to the Earth as a function of the heliocentric distance of initial agglomeration. While this step encompasses a variety of processes such as drift of meter-sized boulders, orbital evolution of parent bodies and eventually ejected meteoroids, we will be content in assuming a \textit{uniform} probability of delivery of material to Earth up to a maximum initial heliocentric distance $R_{\rm max}$. In other words, the distribution we are calculating will be representative of water ice accreted inside $R_{\rm max}$. If the asteroid main belt accreted \textit{in situ}, $R_{\rm max}$ could be taken to correspond to its outer edge, near 3 AU. If, on the other hand, significant redistribution of planetesimals occurred, e.g. during the ``Grand Tack'' studied by \citet{Walshetal2011}, $R_{\rm max}$ could be considerably larger.% (i.e. meteorite parent bodies would represent a wider range of formation locations than the present-day main belt).  

%By excluding (water-poor and therefore less studied) OCs, we avoid sampling biases as well as some main belt/Earth transport biases. 

\section{Results}
\label{Results}

\begin{figure}
\resizebox{\hsize}{!}{
\includegraphics{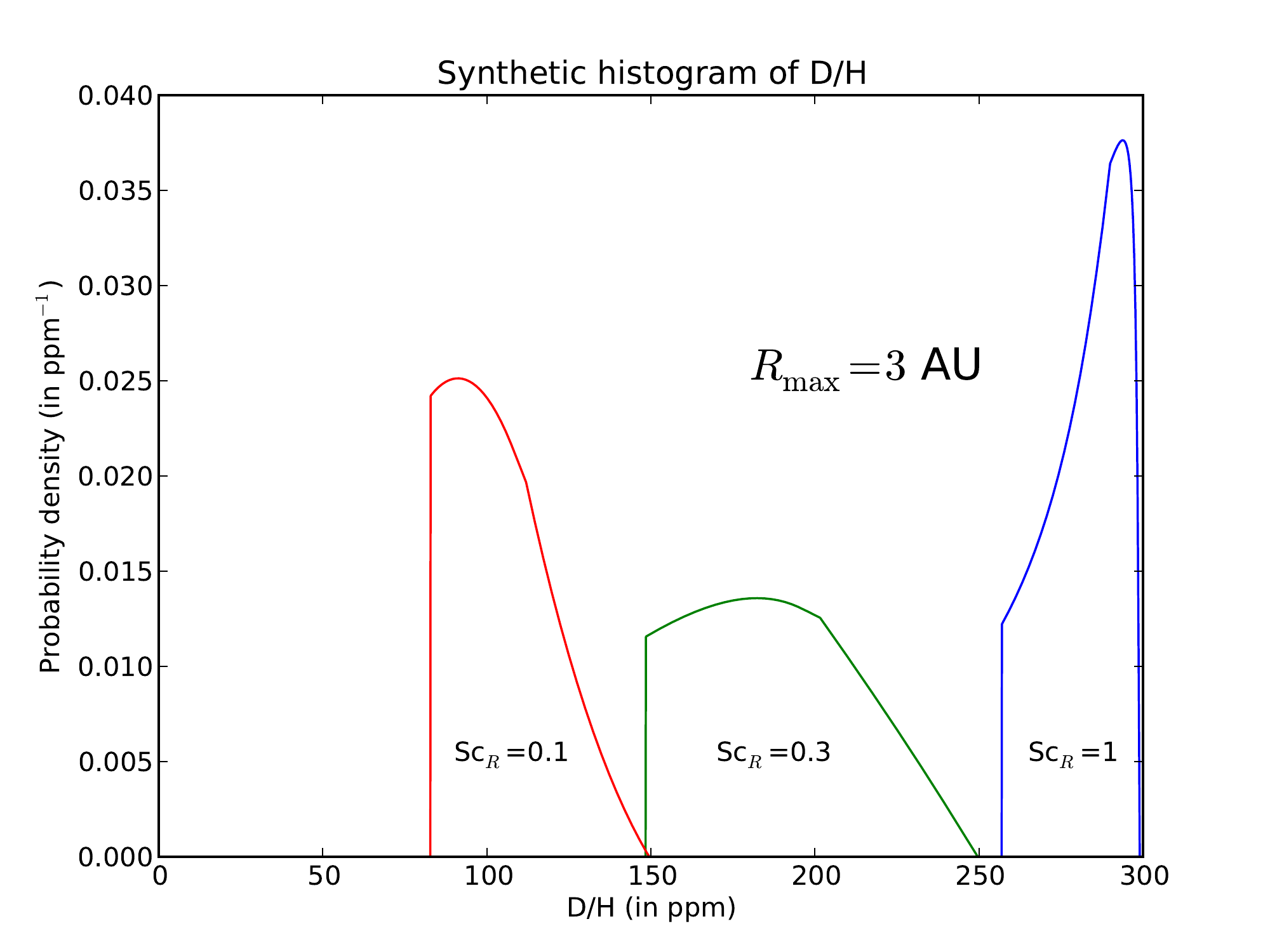}
}
\resizebox{\hsize}{!}{
\includegraphics{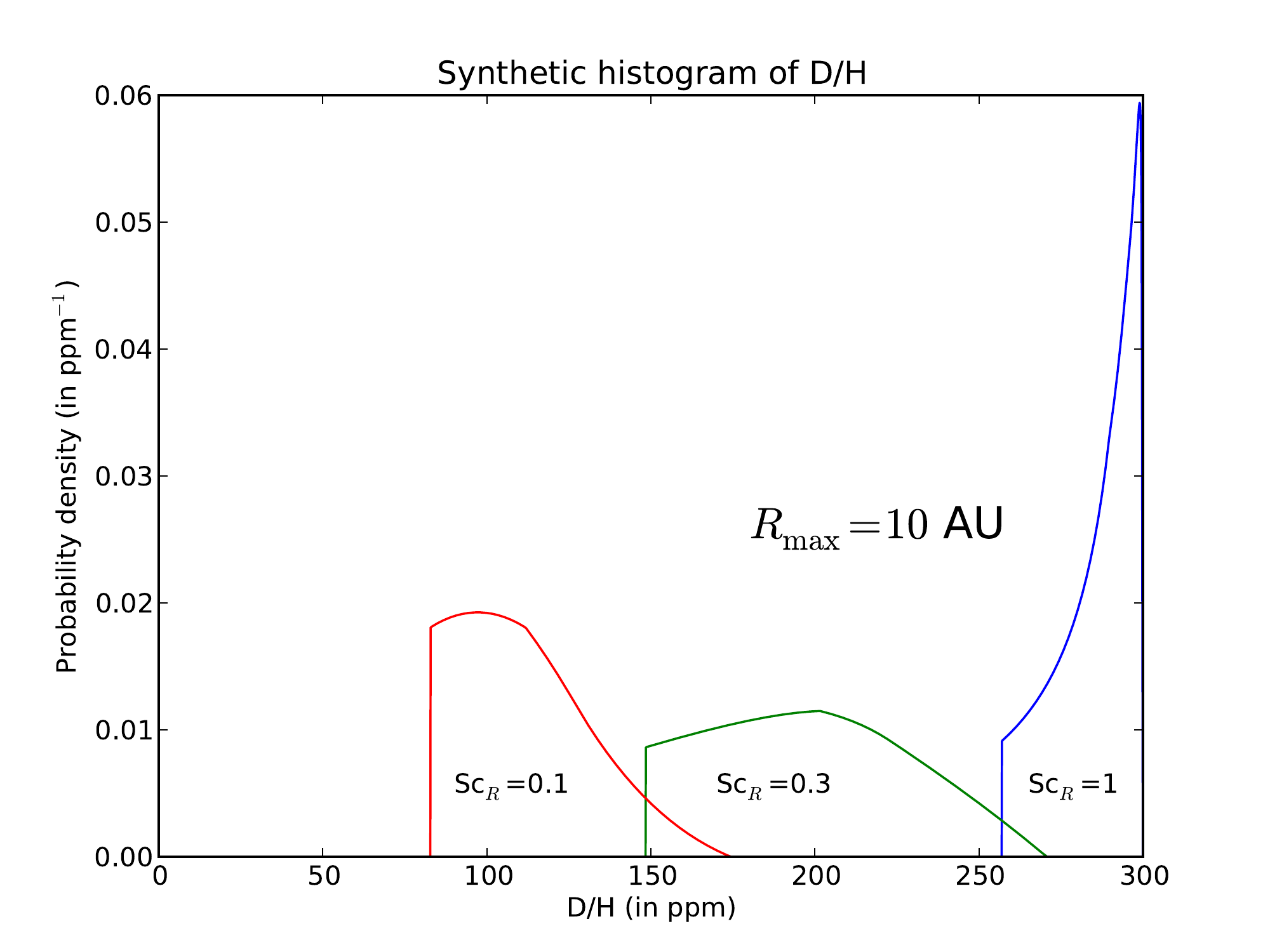}
}
\caption{Probability distribution function of D/H calculated for different values of the radial Schmidt number Sc$_R$, assuming a maximum heliocentric distance of accretion $R_{\rm max}$ of 3 AU (Top) and 10 AU (Bottom). The lower the Sc$_R$ (that is, the larger the diffusivity), the more the PDF is peaked toward low values, while the larger the $R_{\rm max}$, the more the PDF is skewed toward heavy isotopic compositions.}
\label{PDF}
\end{figure}

\begin{figure}
\resizebox{\hsize}{!}{
\includegraphics{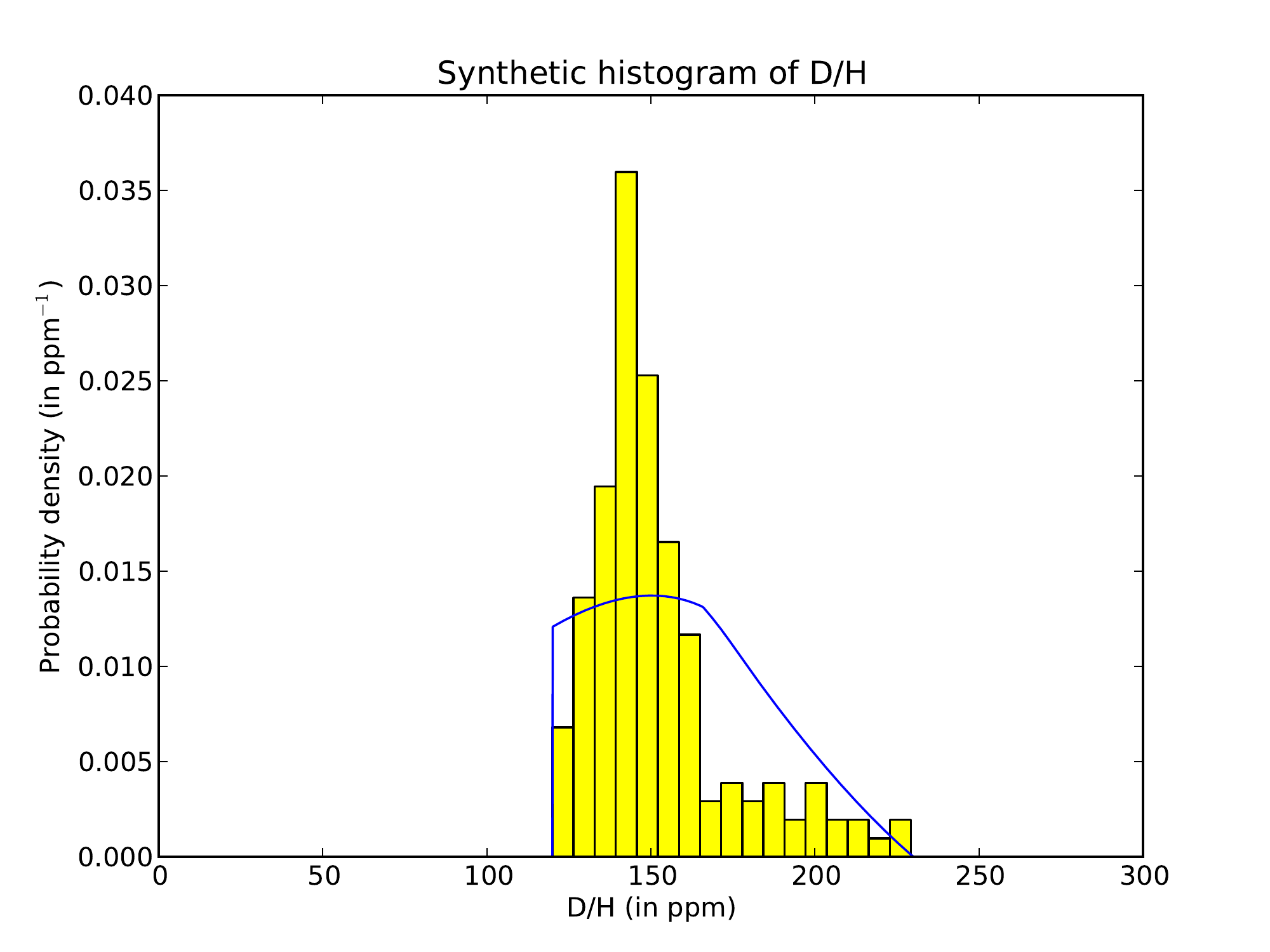}
}
\caption{Probability distribution function of D/H values in bulk carbonaceous chondrites (CI, CO, CV, CM) compared to the theoretical PDF adjusted to fit the minimum and maximum of the distribution. The comparison is warranted only if isotopic exchange with organic matter on the parent body is negligible. Data from \citet{Kerridge1985, Kolodnyetal1980, Boato1954, RobertEpstein1982, McNaughtonetal1982, YangEpstein1983, Pearsonetal2001, Alexanderetal2012}.}
\label{histogram}
\end{figure}

  Under the above hypotheses, meteoritic water, having been accreted from a range of heliocentric distances (between the snow line and $R_{\rm max}$) and at various times, will necessarily exhibit a range of D/H ratios and one can define a probability distribution function (PDF) for that quantity. The derivation and the expression of the PDF are presented in \ref{detailed calculation} and plots of it are shown in Fig. \ref{PDF}. It is notable that they are independent of $\alpha$%\footnote{It must be kept in mind that our calculation still depends on the assumption of \textit{constant} $\alpha$. However, the minimum value of the distribution, of most interest in the discussion, is only weakly dependent on this because of the weak insentivity of the viscous temperature to $\alpha$ (see equation (\ref{T steady})).}
, $t_0$, $t_{\rm coag}$(1 AU) $\epsilon_{\rm H_2O}$ and $\kappa$ and depend on $f_T$, $T_{\rm cond}$, $T_{\rm reac}$, (D/H)$_h$, (D/H)$_l$, $R_{\rm max}$ and Sc$_R$, of which only the latter two are considered free parameters. 

  Distributions exhibit a minimum cutoff value (D/H)$_{\rm min,0}$ (see equation (\ref{xmin0})). %, whose sharpness is fairly insensitive to variations of alpha
This minimum value of the distribution stems from the constraint that the accreted water be condensed, and corresponds to the hydrogen isotopic composition of water at the snow line. It is thus governed by isotope exchange kinetics (which determines the composition and origin of equilibrated water) \textit{and} radial transport to the snow line (which determines the proportion of equilibrated water there) alike. It is noteworthy that the D/H ratio at the snow line does not evolve with time (so long it is in the viscous-heating dominated temperature region), as it only depends on the ratio $T_{\rm cond}/T_{\rm reac}$ (see equation (\ref{xmin})). 

As is apparent on each panel of Fig. \ref{PDF}, low values of (D/H)$_{\rm min,0}$ %(relative to the pristine, ``cometary'' value (D/H)$_h$)
 are associated with efficient radial mixing (enabling a high proportion of equilibrated water at the snow line), that is, a small radial Schmidt number, through the relationship (using equation (\ref{xmin0})):
\begin{equation}
\textrm{Sc}_R=\frac{3}{5}\frac{\mathrm{ln}\left(\frac{(\mathrm{D}/\mathrm{H})_h-(\mathrm{D}/\mathrm{H})_l}{(\mathrm{D}/\mathrm{H})_h-(\mathrm{D}/\mathrm{H})_{\rm min,0}}\right)}{\mathrm{ln}\left(\frac{T_{\rm reac}}{T_{\rm cond}}\right)}
\label{ScR from xmin0}
\end{equation}
For our fiducial parameters, if we adopt (D/H)$_{\rm min,0}=120\times 10^{-6}$ from the observed PDF of D/H in carbonaceous chondrites, we obtain Sc$_R$=0.2.

  A low value of Sc$_R$ is also needed if one is to recover the positive skewness of the observed PDF (with a negative slope over most of the range of D/H values). Indeed, equation (\ref{PDF without time dependence}) indicates that the PDF has an overall dependence in $\left[(\mathrm{D}/\mathrm{H})_h-(\mathrm{D}/\mathrm{H})\right]^{1/(3\textrm{\scriptsize Sc}_R)-1}$, notwithstanding ``second-order'' modifications imposed by the constraint $R_{\rm cond}<R<R_{\rm max}$ and the changes in the temperature regime% (from irradiation- to viscous heating-dominated)
. This imposes an overall decreasing trend if Sc$_R\lesssim 1/3$.\footnote{Note that this inequality does not depend on the prescription we have adopted for $\dot{M}(t)$ but does depend on the power law for the irradiation-dominated temperature regime and that of the coagulation timescale: for $T\propto R^{-q}$ and $t_{\rm coag}\propto R^a$, the inequality becomes Sc$_R\lesssim (2(a-q)-1)/3$ (going back to equation (\ref{f(x)}))}
Efficient transport indeed means that the D/H ratio will be close to that at the snow line for a significant fraction of the outer disk, hence the dominance of this value in the PDF.
%Also depends on Rmax

  By comparing the two panels of Fig. \ref{PDF}, it is also apparent that the PDF is also more skewed toward higher D/H ratio if $R_{\rm max}$ increases, as is expected from the monotonic increase of D/H with increasing heliocentric distance in our model. The maximum value (D/H)$_{\rm max,f}$ (see equation (\ref{xmaxf})) is related to $R_{\rm max}$ through the relationship (injecting equation (\ref{ScR from xmin0}) into equation (\ref{xmaxf})):
\begin{equation}
R_{\rm max, AU}=\left(\frac{f_TT_0}{T_{\rm reac}}\right)^2\mathrm{exp}\left(\frac{10}{9}\frac{\mathrm{ln}\left(\frac{T_{\rm reac}}{T_{\rm cond}}\right)\mathrm{ln}\left(\frac{(\mathrm{D}/\mathrm{H})_h-(\mathrm{D}/\mathrm{H})_l}{(\mathrm{D}/\mathrm{H})_h-(\mathrm{D}/\mathrm{H})_{\rm max,f}}\right)}
{\mathrm{ln}\left(\frac{(\mathrm{D}/\mathrm{H})_h-(\mathrm{D}/\mathrm{H})_l}{(\mathrm{D}/\mathrm{H})_h-(\mathrm{D}/\mathrm{H})_{\rm min,0}}\right)}\right)
\label{Rmax evaluated}
\end{equation}
If we adopt (D/H)$_{\rm max,f}=230\times 10^{-6}$, one obtains $R_{\rm max}=6$ AU%; for (D/H)$_{\rm max, f}>260\times 10^{-6}$ (as for micrometeorites), $R_{\rm max}>30$ AU
. $R_{\rm max}$ as expressed above is however quite sensitive to the assumed parameters% (and the somewhat arbitrary termination of the integration)
, and as such our calculation would not conclusively discriminate between \textit{in situ} formation of the asteroid main belt and widespread redistribution \citep{Walshetal2011}.  Note that the evaluations of both Sc$_R$ and $R_{\rm max}$ in equations (\ref{ScR from xmin0}) and (\ref{Rmax evaluated}) are independent of our prescriptions of accretion of solids or the functional form of $\dot{M}(t)$.

  In Fig. \ref{histogram}, we have plotted the theoretical PDF with the Sc$_R$ and $R_{\rm max}$ evaluated above (which by construction adjust the PDF to the minimum and maximum observed values and the histogram of the latter (observed) data). The shape of the observed PDF is reproduced qualitatively (though not quantitatively), with (i) a peak near the lower end of the distribution and (ii) a tail toward high D/H ratios. The peak corresponds to D/H ratios near the snow line, which dominate the distribution because accretion is most efficient there (because of shorter dynamical timescales and higher surface densities) than further from the Sun and also because efficient outward transport has almost homogenized the D/H ratios to the snow line value over an extensive region beyond it. The tail corresponds to water ice accreted some distance beyond the snow line, until the maximum heliocentric radius sampled.

  However, the match is not quantitative, as the observed PDF is more peaked (around D/H = $150\times 10^{-6}$) than the theoretical prediction. Certainly, the simplifications used in the model (e.g. constant $\alpha$, prescription of delivery to Earth etc.) prevent it from yielding realistic PDFs and only allow a proof-of-concept use. We shall now discuss implications of these results as well as assess the limitations of our treatment. %Possible (but non unique) ways to improve the match would be to prescribe a smoother cutoff of the delivery probability to Earth (Section \ref{Accretion and delivery to Earth of meteoritic water}) starting at a shorter heliocentric distance than the calculated $R_{\rm max}$, or to invoke a steeper surface density profile than the constant-$\alpha$, quasi-static disk model used here. 

\section{Discussion}
\label{Discussion}

\subsection{Implications}
\label{Implications}

  From the above results, it appears that the distribution of D/H ratios of carbonaceous chondrites can be accounted for in a scenario of isotopic exchange with hydrogen gas, with cometary D/H values prevailing at large heliocentric distances, under the condition that the radial Schmidt number Sc$_R$ be small ($\lesssim 0.3$). The peak value of the carbonaceous chondrite population would be essentially dictated by the isotopic composition of water near the snow line, where accretion of \textit{condensed} water would be most efficient. %There is also an indication that the range of heliocentric distances of formation of the carbonaceous chondrites parent bodies extended considerably beyond the outer edge of the present-day main-belt or even the current orbit of Jupiter but uncertainties in the chosen parameters make this a less definite statement.

  The low values inferred for Sc$_R$ would be consistent with hydrodynamical turbulence \citep[e.g.][]{Prinn1990,DubrulleFrisch1991,Gail2001} --- e.g. the 0.176 value measured by \citet{Lathropetal1992}, although more experimental (and numerical) data would be desirable. It would not be consistent, however, with magnetohydrodynamic (MHD) turbulence %, except for the lowest values of $\alpha$
 \citep{Johansenetal2006}. MHD turbulence, mainly driven by the magnetorotational instability (MRI; \citealt{BalbusHawley1998}), is widely believed to be the driver of transport in accretion disks. However, there must be a region in the disk, referred to as the \textit{dead zone}, which is too cold and dense for the MRI to operate \citep{Gammie1996}. In the dead zone, turbulence should be of low intensity and hydrodynamical in nature. Thus our results are consistent with the idea that the planetesimals sampled by chondrites formed in this purported dead zone. In fact, ionization fraction calculations have shown that the dead zone would encompass the whole planet-forming region (from $\sim$0.1 to a few 10s of AU, e.g. \citealt{BaiGoodman2009}) and indeed a dead zone would be a most favorable environment for planet formation \citep{Terquem2008}. Moreover the existence of a dead zone has been found to enable the preservation of chondrite components for a few Ma in the disk \citep{Jacquetetal2011a}. 

  The low value of Sc$_R$ would not only have enhanced outward transport of equilibrated water, but also that of higher-temperature components as well (see also \citealt{BockeleeMorvanetal2002}). This could account for the high-temperature minerals, including calcium-aluminum-rich inclusions (CAIs) and chondrule fragments identified in comet Wild 2 samples \citep{Zolenskyetal2006,Bridgesetal2012} and also the relatively high abundances of CAIs in carbonaceous chondrites (see \citealt{Jacquetetal2012S}). %In fact, the problem of outward transport of CAIs in the model here is the same as that with water. Using the same formalism as before (with CAIs playing the role of equilibrated water), the abundance of refractory inclusions at the snow line normalized to that in the CAI-forming region would be $(T_{\rm cond}/T_{\rm CAI})^{5\textrm{\scriptsize Sc}_R/3}$ with $T_{\rm CAI}$ a typical CAI-forming temperature. If we take $T_{\rm CAI}=$1500 K and the Sc$_R$=0.2 estimated above, one obtains 0.5, consistent with reported CAI abundances in carbonaceous chondrites (see \citet{Jacquetetal2012S}).

  If mixing was as efficient as inferred above, the D/H ratio could have been significantly lower than (D/H)$_h$ even at a few tens of AUs, depending on time (see Fig. \ref{D/H vs R}). Thus, comets may have formed with D/H ratios lower than (D/H)$_h$ (the limit at large heliocentric distances), which would account for the emerging range in measured D/H ratios of cometary water \citep{Hartoghetal2011, BockeleeMorvanetal2012}, consistent with the idea of an asteroid-comet continuum \citep{Gounelleetal2008, Brianietal2011}. Even the $300\times 10^{-6}$ value adopted for (D/H)$_h$ on the basis of measurements of known comets could then actually be a lower bound of its real value.

\subsection{Modeling caveats}
\label{Caveats}

\begin{figure}
\resizebox{\hsize}{!}{
\includegraphics{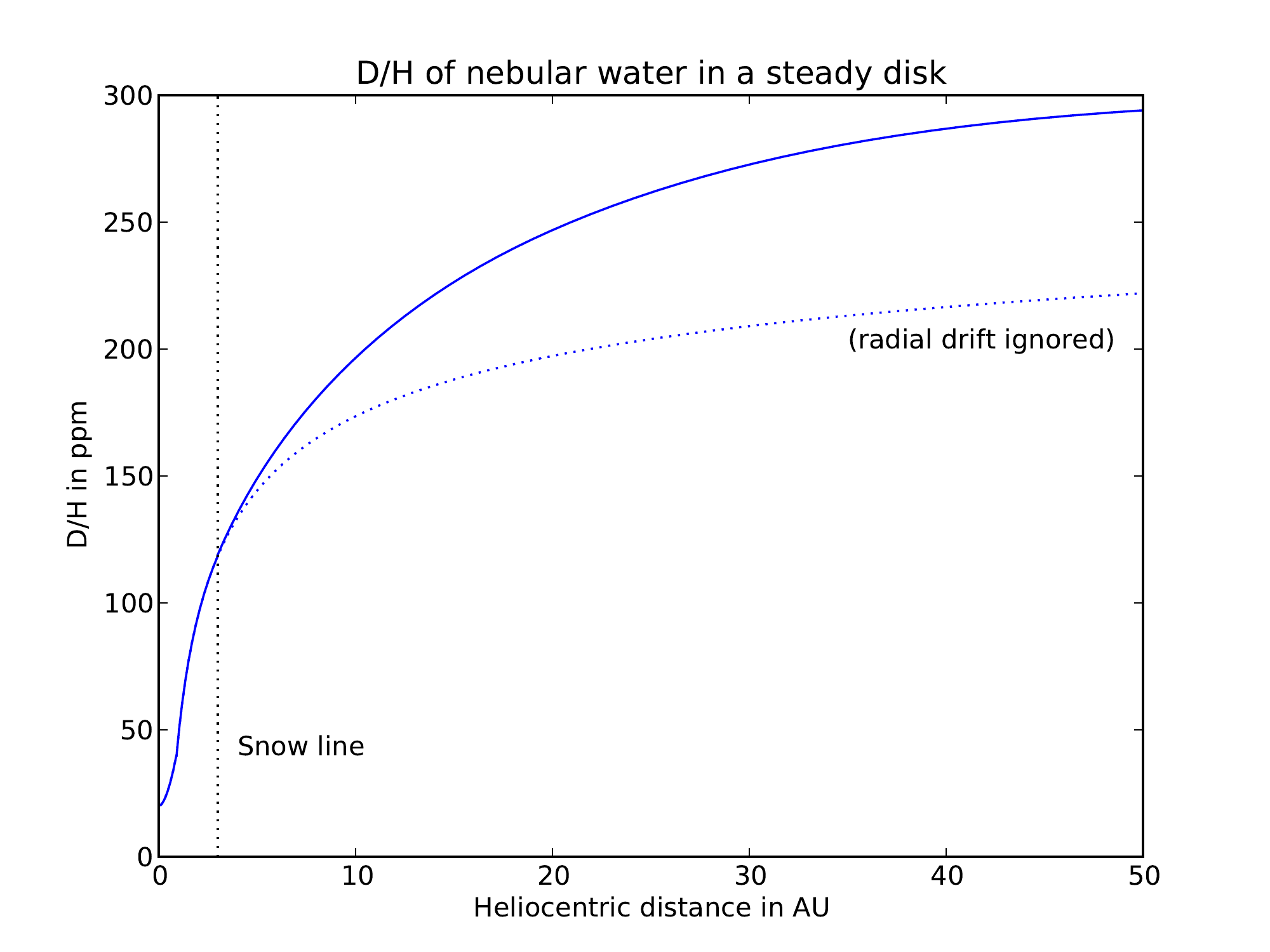}
}
\caption{The D/H ratio as a function of heliocentric distance of accretion, for the radial Schmidt number inferred from this study (0.2) at a fiducial time where the snow line lies at 3 AU from the Sun. We have taken $\kappa=0.5\:\mathrm{m^2/kg}$ and $\alpha=10^{-4}$ (as appropriate in a dead zone). To realistically extrapolate to the comet-forming regions, we have here taken into account the effect of radial drift by considering millimeter-sized grains ($\rho_sa=1\:\mathrm{kg/m^2}$ ; see \ref{appendX}). The D/H curve without this effect is also shown for comparison (dotted line).} %$\dot{M}=3\times 10^{-9}\:M_\odot.a^{-1}$
\label{D/H vs R drift}
\end{figure}

  Of course the validity of the conclusions of this study can only be as good as that of its underlying assumptions. 

  First, the quasi-static approximation we have used requires, as we recall from Section \ref{The disk}, that the viscous timescale $t_{\rm vis}(R)$ be short relative to the time of chondrite accretion and that of evolution of the mass accretion rate. In a dead zone, whose existence we have inferred above, $\alpha$ would likely be small, around $10^{-4}$ \citep[e.g.][]{FlemingStone2003,IlgnerNelson2008,OishiMcLow2009,Turneretal2010}, so that $t_{\rm vis}$ would be longer than our nominal evaluation in equation (\ref{tvis}), but at $R=3$ AU, it would be $\sim$1 Ma, still shorter than the accretion time of chondrites after the start of the solar system ($\sim$1-5 Ma, see e.g. \citet{Villeneuveetal2009,Connellyetal2012})%, although our approximation would break down for $R\gtrsim$ 10 AU, but this would be outside our regions of interest.
. Admittedly, the timescale constraint would be quite marginally satisfied, so that, conceivably, future re-examination of this problem in time-dependent simulations of disks with dead zones, with self-consistently evaluated $\alpha$ (see e.g. \citet{Zhuetal2010b}), could unveil new interesting effects. Nonetheless, this would not affect our conclusion that MRI-turbulent disk models, with higher $\alpha$---hence largely satisfying the timescale constraint--- and Sc$_R$, cannot reproduce the D/H ratio of chondrites and our inference of the presence of dead zone thus appears robust in this respect. %constant alpha vs. outer edge of dead zone far away.

  Our steady-state disk model neglects photoevaporation during the bulk of the disk's lifetime%, generally thought to be important in late stages of disk evolution
. However, if, as argued by \citet{Desch2007}, the disk was subject to intense photoevaporation from the outset, the net gas flow could have been \textit{outward}, and equilibrated water would have reached the snow line almost undiluted; i.e. water at the snow line would have a D/H ratio close to (D/H)$_l\approx 40\times 10^{-6}$. If such isotopically light \textit{bulk} preaccretionary water compositions were to be found in the future, this would be a possibility worth investigating ; however, to date, such isotopic compositions have not been measured.

% As to a proper treatment of reaction kinetics, beyond a smoothing of the lower end of the distribution, the effect would likely be small: one correction would be to account for a secular decrease of $T_{\rm reac}$ (see equation (\ref{teq})) by a few tens of degrees at most, but this would be accompanied by an increase in $x_l$ and thus would weakly affect equation (\ref{ScR from xmin0}). 

  In this work, nebular water has been assumed to be dynamically coupled to the gas. Depending on the fragmentation and the bouncing barriers \citep[e.g.][]{Zsometal2010,Birnstieletal2012}, it is however conceivable that ice-bearing particles grow to millimeter-size and beyond so that drift due to gas drag would be significant (see equation (\ref{SR steady})) before they are incorporated in planetesimals. This would likely lead to significant enhancements of water abundance inside and at the snow line, and depletions further outwards \citep[e.g.][]{CuzziZahnle2004,CieslaCuzzi2006}. From equation (\ref{eq sur tot avec S}) derived in \ref{appendX}, it can be seen that the fraction of equilibrated water would be lower, and thus D/H higher, than under our tight coupling calculation, because of increased import of ``cometary water''. One example of this effect is shown in Fig. \ref{D/H vs R drift} using the parameters inferred from this study. While the effect might be limited for the carbonaceous chondrites which we have focused on---for which \citet{Jacquetetal2012S} inferred that millimeter-sized components were not significantly drifting---, it would certainly accelerate the convergence of the D/H ratio toward its asymptotic value in the comet-forming region (for $S_R>1$, see \ref{appendX}). This could account for the apparent carbonaceous chondrite/comet hydrogen isotopic dichotomy, in the sense that relatively little material with intermediate composition would then exist (but does exist, as emphasized at the end of Section \ref{Implications}). However, regardless of the actual quantitative importance of radial drift, it must be noted that
%, which would stretch the high-D/H tail of the PDF to larger values for a given $R_{\rm max}$ (which is anyway weakly constrained by our model). However, while this would stretch the high end of the PDF to heavier values,
 inasmuch as nebular water \textit{inside} the snow line would be in vapor state --- even meter-sized solids would lose their water by vaporization before they can travel much inward of it \citep{CuzziZahnle2004} --- and thus continue to be tightly coupled to the gas, the transport of equilibrated water until the snow line would be unchanged (see equation \ref{eq sur tot avec S}) --- save for some transition period preceding establishment of the quasistatic regime. Therefore, our estimate of Sc$_R$ from the minimum D/H of the distribution (established at the snow line) would be unaffected and is thus robust in this regard too.

  We have also argued that accretion was inefficient, allowing us to ignore feedback on the transport of nebular water. If this approximation were not to hold, rapid accretion of planetesimals at the snow line could progressively deplete the inner solar system in water \citep[e.g.][]{CieslaCuzzi2006}. Although, again, this would likely not affect the D/H ratio at the snow line, the effects on the shape of the PDF have yet to be determined by dedicated numerical simulations including both coagulation and shattering \citep[see][]{Yangetal2012}.

%Including the effect of coagulation on the water/gas ratio would lead to a marked drop (for relatively rapid coagulation timescales as mentioned earlier) of this ratio just outside the snow line \citep{CieslaCuzzi2006} and thus would \textit{diminish} the proportion of equilibrated water, such that even smaller value of Sc$_R$ would be indicated%, hereby not endangering our conclusion on the hydrodynamical nature of turbulence. Likewise: shattering => weakened dependence on $\Sigma$

\subsection{On the interpretation of D/H chondrite data}

  In this work, we have compared our PDF of the isotopic composition of water to bulk chondrite compositions (see Fig. \ref{histogram}), and, in doing so, implicitly assumed that the latter reflected the composition of pre-accretionary water now locked in hydrated silicates. However, hydrated silicates are not the only contributors to the hydrogen budget of carbonaceous chondrites, as the latter also contain organic matter. The rationale for making this identification nonetheless is that inasmuch organic matter does not account for more than 10 \% of the hydrogen, at their \textit{presently measured} D/H ratios (mostly $\lesssim 400\times 10^{-6}$---still excluding CR chondrites---; \citet{Alexanderetal2010}), the incurred shifts to the isotopic composition would be small ($\lesssim 20\times 10^{-6}$ ; \citet{Robert2006}) compared to the observed range of D/H ratios. However, this ignores the possibility of isotopic exchange between organic matter and water on the parent body, in which case the organic matter could have been much more D-rich initially than presently measured. In fact, \citet{Alexanderetal2012} showed that the D/H ratio of bulk carbonaceous chondrites correlated positively with the C/H ratio, and interpreting this as a mixing trend, extrapolated an initial D/H ratio for organic matter comparable to that of CR chondrites ($\approx 700\times 10^{-6}$), and consequently, a lighter isotopic composition for pre-accretionary water, e.g. $(86.5\pm 3.5)\times 10^{-6}$ for CM chondrites\footnote{This would incidentally explain some low-D/H point measurements in LL3 chondrites \citep{Delouleetal1998}.}. In that case, while (direct) comparison of the theoretical PDF with the histogram plotted in Fig. \ref{histogram} would no longer make sense, (D/H)$_{\rm min,0}$ would be constrained to be lower than this latter value, and our formalism would thus constrain Sc$_R$ to be $\lesssim 0.1$ (see equation (\ref{ScR from xmin0})), that is, a yet more efficient radial diffusion would be indicated. %would explain also clustering to a single intercept (but trend could be heteregeneous accretion, Alexander et al 2012 MetSoc)

  It must be cautioned, though, that \textit{in situ} observations of CI, CM and CR chondrite matrices have revealed little evidence of isotopic exchange between the isotopically heterogeneous organic matter and the intermixed hydrous minerals \citep{Remusatetal2010} and organic matter in the least aqueously altered CM chondrite Paris has an hydrogen isotopic signature indistinguishable of that of the other CM \citep{Remusatetal2011}. One possible explanation for the D/H - C/H correlation of \citet{Alexanderetal2012} could be that the cometary water endmember was somehow coupled to D-rich organic matter in the disk, consistent with the high carbon contents of cometary grains \citep[e.g.][]{Wozniakiewiczetal2012}. In that case, the mixing trend would not actually extrapolate to the composition of the organic matter but to that of the resulting \textit{composite} (water+organics) C- and D-rich endmember. In this picture, %(though not require)
 the hydrogen isotopic signatures of organic matter and hydrated silicates could be actually largely pristine, as inferred by \citet{Remusatetal2010}. Whatever that may be, and whether D/H ratios of bulk carbonaceous chondrites truly reflect the D/H ratios of preaccretionary water or only provide \textit{upper} limits, we stress that low Schmidt numbers are robustly required in the framework of our formalism.

  %---possibly explaining the narrowness of the observed (extrapolated) peak

% It must be cautioned, though, that the proposed corrections assume negligible water loss from the system since accretion, whereas oxygen isotopic systematics generally suggest relatively high initial water/rock ratios (0.1-0.3 by mass in CM chondrites according to the \citealt{ClaytonMayeda1999} model) compared to the currently measured ones.

 While the present work has mainly focused on carbonaceous chondrites, it is noteworthy that non-carbonaceous chondrites are generally richer in deuterium than the former \citep[e.g.][]{Robert2006,Alexanderetal2012}\footnote{\citet{Alexanderetal2012MetSoc} suggest that the high D/H ratios of these meteorites could be due to oxidation of metallic iron by water producing isotopically light hydrogen gas escaping from the system. However, while this may have affected metamorphosed chondrites \citep{Alexanderetal2010,McCantaetal2008}, the most unequilibrated ordinary chondrites also show these high values, with considerable heterogeneity consistent with the pristinity of this signature \citep{Piani2012}.}. As the D/H ratio of nebular water increases with time in our scenario, this could imply that they accreted \textit{later} than the former, as suggested by \citet{Jacquetetal2012S} based on the modeled redistribution of chondrite components in the disk. Conceivably, radial drift, as alluded to in the preceding subsection, which would be most pronounced as the coupling of the grains with the less dense gas would be looser then, might have contributed to significant enhancements of the D/H ratios for them. In fact, \citet{Jacquetetal2012S} specifically inferred that millimeter-size components were significantly drifting for these chondrites, and from equation (\ref{eq sur tot avec S}), the nebular water should be isotopically close to cometary values, which appears to be the case for clays from unequilibrated ordinary chondrites \citep[e.g.][]{Robert2006,Alexanderetal2012}. 

  A later formation of the non-carbonaceous chondrites would avoid the alternative possibility of an accretion \textit{further} from the Sun than carbonaceous chondrites, which would run counter to the observed distribution of asteroid classes %generally associated to these respective categories
\citep[e.g.][]{Burbineetal2008}%Supercometary D/H ratios -> xh evolves input of more pristine interstellar matter (see IOM for OCs, Aleon 2010)
. A later formation time could also account for the high D/H values of CR chondrites too---consistent with their young chondrule Al-Mg ages \citep{KitaUshikubo2012}---, which could mark an intermediate status between the other carbonaceous and the non-carbonaceous chondrites---along with their rather low refractory inclusions abundance, their limited refractory element enrichment relative to Mg, and their relatively heavy oxygen isotopic composition \citep{ScottKrot2003}.

\section{Summary}
\label{Conclusion}

We have considered a simplified analytic model for water transport and isotopic evolution in an evolving disk based on the following main assumptions:
\begin{itemize}
\item[(i)] The regions of interest of the disk are approximated by a stationary model, with a constant turbulence parameter $\alpha$.
\item[(ii)] Accretion is inefficient and does not significantly affect the transport of nebular water.
\item[(iii)] Water far from the Sun is assumed to have the same D/H ratio as cometary water, and any batch of water having experienced temperatures above a ``reaction temperature'' $T_{\rm reac}$ has had its D/H reset to a fixed value (D/H)$_l$.
\item[(iv)] Accretion is modeled by a coagulation timescale scaling with the orbital period and water is incorporated as ice only. A uniform probability of delivery to Earth is applied to agglomeration locations up to a maximum heliocentric distance $R_{\rm max}$.
\end{itemize}
  In this model, the D/H ratio of water is an increasing function of time and heliocentric distance. After integration over time, the probability distribution function (PDF) of the D/H ratio in accreted water is found to depend essentially on $R_{\rm max}$ and (most sensitively) on the radial Schmidt number Sc$_R\equiv \alpha/\delta_R$ with $\delta_R$ parameterizing the diffusivity. The minimum cutoff of the PDF is determined by isotopic composition of water at the snow line, while the isotopically heavy tail is dictated by $R_{\rm max}$. 

  It appears that the model is able to broadly account for the observed PDF in carbonaceous chondrites if low values of Sc$_R$ (around 0.1-0.3)---i.e. efficient outward diffusion---are assumed in order to reproduce the low D/H values of most chondrites and the positive skewness of the observed distribution. This would be most consistent with hydrodynamical turbulence as expected to prevail in the dead zone of the protoplanetary disk. Efficient outward diffusion would also have enabled the transport of high-temperature minerals to comets. The high D/H ratios in CR chondrites and non-carbonaceous chondrites could indicate an accretion later than most carbonaceous chondrites. %In the self-shielding scenario, this would be consistent with the heavier oxygen isotopic composition of the former.

  The effects of radial drift and higher accretion efficiencies on the transport of water and its hydrogen isotopic composition have yet to be investigated.
  
\section*{Acknowledgement}
We thank the two anonymous referees for their reviews which improved the clarity of the paper in particular as to the effects of radial drift and the transition to cometary isotopic signatures.

\begin{appendix}

\section{Diffusion of equilibrated water in disks}% Interdiffusion of components in disks}
\label{appendX}

In this appendix, we calculate the steady-state profile of nebular water concentration and its equilibrated fraction. To allow discussion of the tight coupling assumption in Section \ref{Discussion}, we take into account the finite size of ice-bearing particles beyond the snow line, and thence the effects of gas drag, so that their velocity becomes, ignoring any feedback of the solids on the gas \citep[e.g.][]{Birnstieletal2010}:
\begin{equation}
v_R=%\equiv u_R + v_{\rm drift}\equiv  
\frac{1}{1+\textrm{St}^2}\left(-\frac{3}{\Sigma R^{1/2}}\frac{\partial}{\partial R}\left(R^{1/2}\Sigma\nu\right) + \frac{\tau}{\rho}\frac{\partial P}{\partial R}\right),
\label{vR}
\end{equation}
with $\nu=\alpha c_s^2/\Omega$ the turbulent viscosity, %$v_{\rm drift}$ the drift velocity due to gas drag, 
$\tau$ the %corresponding
 gas drag stopping time, $\textrm{St}\equiv\Omega\tau$ the Stokes number, %(assumed to be $< \Omega^{-1})$ %\footnote{Strictly speaking, equation (\ref{vR}) assumes that $\tau \ll \Omega^{-1}$, otherwise, an overall factor of $(1+(\Omega\tau)^2)^{-1}$ would appear \citep[e.g.][]{Birnstieletal2010}, but it would be cancelled by a similar factor that would modify the diffusion coefficient \citep{YoudinLithwick2007} in the ratio $v_R/D_R$ but still appear in the integrand} (assumed to be $\ll \Omega^{-1}$)
 and $P$ and $\rho$ the gas pressure and density, respectively.

  In steady state, the mass accretion rate of nebular water
\begin{equation}
\dot{M}_{\rm H_2O}=2\pi R\left(-\Sigma_{\rm H_2O}v_R+D_R\Sigma\frac{\partial}{\partial R}\left(\frac{\Sigma_{\rm H_2O}}{\Sigma}\right)\right),
\label{Mdotp}
\end{equation}
is constant. We have introduced the diffusion coefficient modified by finite particle size as \citep{YoudinLithwick2007}
\begin{equation}
D_R=\frac{\delta_Rc_s^2/\Omega}{1+\textrm{St}^2}.
\end{equation}
Equation (\ref{Mdotp}) may be viewed as a first-order ordinary differential equation in $\Sigma_{\rm H_2O}/\Sigma$. It is noteworthy that the corresponding homogeneous equation is the equation governing the transport of equilibrated water, since its concentration vanishes at infinity, so that:
\begin{equation}
\frac{\Sigma_{\rm H_2O, eq}}{\Sigma}\propto \mathrm{exp}\left(\int^R\frac{v_R}{D_R}\mathrm{d}R'\right) \propto \frac{\exp{\left(\int^RS_R\frac{\partial\mathrm{ln}P}{\partial R}\mathrm{d}R'\right)}}{\left(\Sigma\nu R^{1/2}\right)^{3\textrm{\scriptsize Sc}_R}} 
\end{equation}
where the second proportionality relationship assumes that Sc$_R$ is radially constant and we have coined $S_R=\textrm{St}/\delta_R$ which is a measure of gas-grain decoupling \citep{Jacquetetal2012S}. In the Epstein drag regime, for spherical grains of density $\rho_s$ and radius $a$ (averaged over the size distribution), we have:
\begin{eqnarray}
S_R=\frac{\pi}{2}\frac{\rho_sa}{\Sigma\delta_R}& = & \frac{3\pi^2\textrm{Sc}_R}{2}\frac{\rho_sa c_s^2}{\dot{M}\Omega}\nonumber\\
& = & 0.1 \frac{\textrm{Sc}_RR_{\rm AU}^{3/2}}{\dot{M}_{-8}}\left(\frac{\rho_sa}{1\:\mathrm{kg/m^2}}\right)\left(\frac{T}{300\:\mathrm{K}}\right),
\label{SR steady}
\end{eqnarray}
where the last two equations pertain to a steady disk as assumed in the main text. Note that the normalizing value $\rho_sa=1\:\mathrm{kg/m^2}$ corresponds to millimeter-sized grains (the typical size of non-matrix chondrite components). 

  By requiring that $\Sigma_{\rm H_2O}/\Sigma$ does not diverge at the disk's inner edge (whose heliocentric distance we denote by $R_{\rm in}$, taken to be zero in the main text), equation (\ref{Mdotp}) may be then integrated as:
\begin{equation}
\frac{\Sigma_{\rm H_2O}}{\Sigma}=\mathrm{exp}\left(\int_{R_{\rm in}}^R\frac{v_R}{D_R}\mathrm{d}R'\right)\int_{R_{\rm in}}^R
\mathrm{exp}\left(-\int_{R_{\rm in}}^{R'}\frac{v_R}{D_R}\mathrm{d}R''\right)\frac{\dot{M}_{\rm H_2O}\mathrm{d}R'}{2\pi R'\Sigma D_R}
\end{equation}
For $S_R\ll 1$ (in particular inside the snow line where $S_R=0$), this is a constant %(as can be seen in the simulations of \citet{CieslaCuzzi2006})%assuming steady disk (only radial transport)
, and for $S_R\gg 1\gg \textrm{St}$, it falls off as $1/S_R$. Then, the fraction of equilibrated water is:
\begin{equation}
\frac{\Sigma_{\rm H_2O, eq}}{\Sigma_{\rm H_2O}} \propto  \left[\int_{R_{\rm in}}^R\mathrm{exp}\left(-\int_{R_{\rm in}}^{R'}\frac{v_R}{D_R}\mathrm{d}R''\right)\frac{\mathrm{d}R'}{R'\Sigma D_R}\right]^{-1}
%& \propto & \left[\int_{R_{\rm in}}^R\left(\Sigma\nu\right)^{3\textrm{\scriptsize Sc}_R-1}R'^{3\textrm{\scriptsize Sc}_R/2-1}\mathrm{exp}\left(-\int^{R'} S_R\frac{\partial\mathrm{ln}P}{\partial R}\mathrm{d}R''\right)\mathrm{d}R'\right]^{-1}\nonumber
\end{equation}
Enforcing that the left-hand-side be unity at $R=R_{\rm reac}$, this becomes:
\begin{eqnarray}
\frac{\Sigma_{\rm H_2O, eq}}{\Sigma_{\rm H_2O}}=\frac{2}{3\textrm{Sc}_R}
\left(\sqrt{R_{\rm reac}}-\sqrt{R_{\rm in}}\right)^{3\textrm{\scriptsize Sc}_R}%\nonumber\\
\Bigg[\int_{R_{\rm in}}^R\left(\sqrt{R'}-\sqrt{R_{\rm in}}\right)^{3\textrm{\scriptsize Sc}_R-1}\nonumber\\
\mathrm{exp}\left(-\int_{R_{\rm reac}}^{R'} S_R\frac{\partial\mathrm{ln}P}{\partial R}\mathrm{d}R''\right)\left(1+\textrm{St}^2\right)\frac{\mathrm{d}R'}{\sqrt{R'}}\Bigg]^{-1}.
\label{eq sur tot avec S}
\end{eqnarray}
For $S_R\ll 1$, this amounts to equation (\ref{X eau}) for $R_{\rm in}\ll R_{\rm reac}<R$.

\section{Kinetics of water-hydrogen isotopic exchange}
\label{kinetics}
Ignoring transport, and given that D/H $\ll 1$ and H$_2$O/H$_2\ll 1$, the equation governing the evolution of D/H of water may be written as \citep{LecluseRobert1994}:
\begin{equation}
\frac{\mathrm{d}}{\mathrm{d}t}\left(\frac{\mathrm{D}}{\mathrm{H}}\right)=k^-(T)n_{\rm H_2}\left(\left(\frac{\mathrm{D}}{\mathrm{H}}\right)_{\rm eq}(T)-\left(\frac{\mathrm{D}}{\mathrm{H}}\right)\right)
\label{dx/dt}
\end{equation}
with $n_{\rm H_2}$ the number density of hydrogen molecules, (D/H)$_{\rm eq}$ the equilibrium values (see e.g. \citealt{Richetetal1977}) and the rate constant
\begin{equation}
k^-(T)=2\times 10^{-28}\:\mathrm{exp}\left(-\frac{5170\:\mathrm{K}}{T}\right) \mathrm{m^3/s}.%Ae^{-T_a/T}
\end{equation}
%with $A=2.2\times 10^{-28}\:\mathrm{m^3/s}$ and $T_a=5170$K \citep{LecluseRobert1994}.

  Given that, in our steady-state model, the midplane number density can be expressed as a function of temperature (assumed to be vertically constant) as 
\begin{eqnarray}
n_{\rm H_2}=\frac{\Sigma\Omega}{\sqrt{2\pi}mc_s}&=&\frac{16}{9}\left(\frac{3}{\dot{M}\alpha}\right)^{1/3}\left(\frac{2T^{11}}{\pi k_B^5m}\right)^{1/6}\left(\frac{\sigma_{\rm SB}}{\kappa}\right)^{2/3},%\nonumber\\
%&=& 8\times 10^{18}\:\mathrm{m^{-3}}\left(\frac{T}{100\:\mathrm{K}}\right)^{11/6}\left(\frac{10^{-3}}{\alpha \dot{M}_{-8}}\right)^{1/3}\left(\frac{0.5\rm \: m^2/kg}{\kappa}\right)^{2/3}
\end{eqnarray}
the characteristic equilibration timescale resulting from equation (\ref{dx/dt}) is:
\begin{eqnarray}
t_{\rm eq} & \equiv & \frac{1}{k^-(T)n_{\rm H_2}}\nonumber\\
&= & 0.1\:\mathrm{Ma}\frac{\mathrm{exp}\left(\frac{5170\:\mathrm{K}}{T}\right)}{T_K^{11/6}}\left(\frac{\alpha \dot{M}_{-8}}{10^{-3}}\right)^{1/3}\left(\frac{\kappa}{0.5\:\rm m^2/kg}\right)^{2/3}.
\label{teq}
\end{eqnarray}
This is a sharply decreasing function of temperature. While, at high temperature, $t_{\rm eq}$ is short compared to the transport timescale $t_{\rm vis}$ so that water vapor and hydrogen gas can be considered in equilibrium, at lower temperature (at larger heliocentric distances), $t_{\rm eq}$ becomes long compared to $t_{\rm vis}$ and isotopic exchange is essentially quenched. The temperature $T_{\rm reac}$ marking the transition between the two regimes can be determined by setting:
\begin{equation}
t_{\rm eq}(T_{\rm reac})\equiv t_{\rm vis}(T_{\rm reac}).
\label{Treac defined}
\end{equation}
 Water originating from inside $R_{\rm reac}$ (``equilibrated water'') will then essentially have the isotopic composition it had when it last equilibrated with hydrogen gas, i.e. at $T=T_{\rm reac}$, that is, its D/H ratio will be (D/H)$_l\equiv$(D/H)$_{\rm eq}(T_{\rm reac})$.

  Numerically, equation (\ref{Treac defined}) can be expressed as:
\begin{equation}
\frac{\mathrm{exp}\left(\frac{5170\:\mathrm{K}}{T_{\rm reac}}\right)}{T_{\rm reac, K}^{5/18}}=\frac{5\times 10^3}{\dot{M}_{-8}^{1/9}}\left(\frac{10^{-3}}{\alpha}\right)^{13/9}\left(\frac{0.5\:\mathrm{m^2/kg}}{\kappa}\right)^{5/9}.
\end{equation}
The left-hand-side being a sharply decreasing function of $T_{\rm reac}$, the sensitivity on $\dot{M}$ is completely negligible and that on $\alpha$ %and the frequency factor
is also fairly weak. We shall thus adopt the solution of this equation for $\alpha = 10^{-3}$, $\dot{M}=10^{-8}\:\mathrm{M_\odot/a}$, $\kappa=0.5\:\mathrm{m^2/kg}$, which is $T_{\rm reac}=500\:\mathrm{K}$.  Adopting a protosolar D/H ratio (for the H$_2$ gas) of $(20\pm 3.5) \times 10^{-6}$ \citep{GeissGloeckler2003}, the fractionation factor given by \citet{Richetetal1977} yields (D/H)$_l=40\times 10^{-6}$.

\section{Calculation of the PDF of D/H}
\label{detailed calculation}

\begin{figure}
\resizebox{\hsize}{!}{
\includegraphics{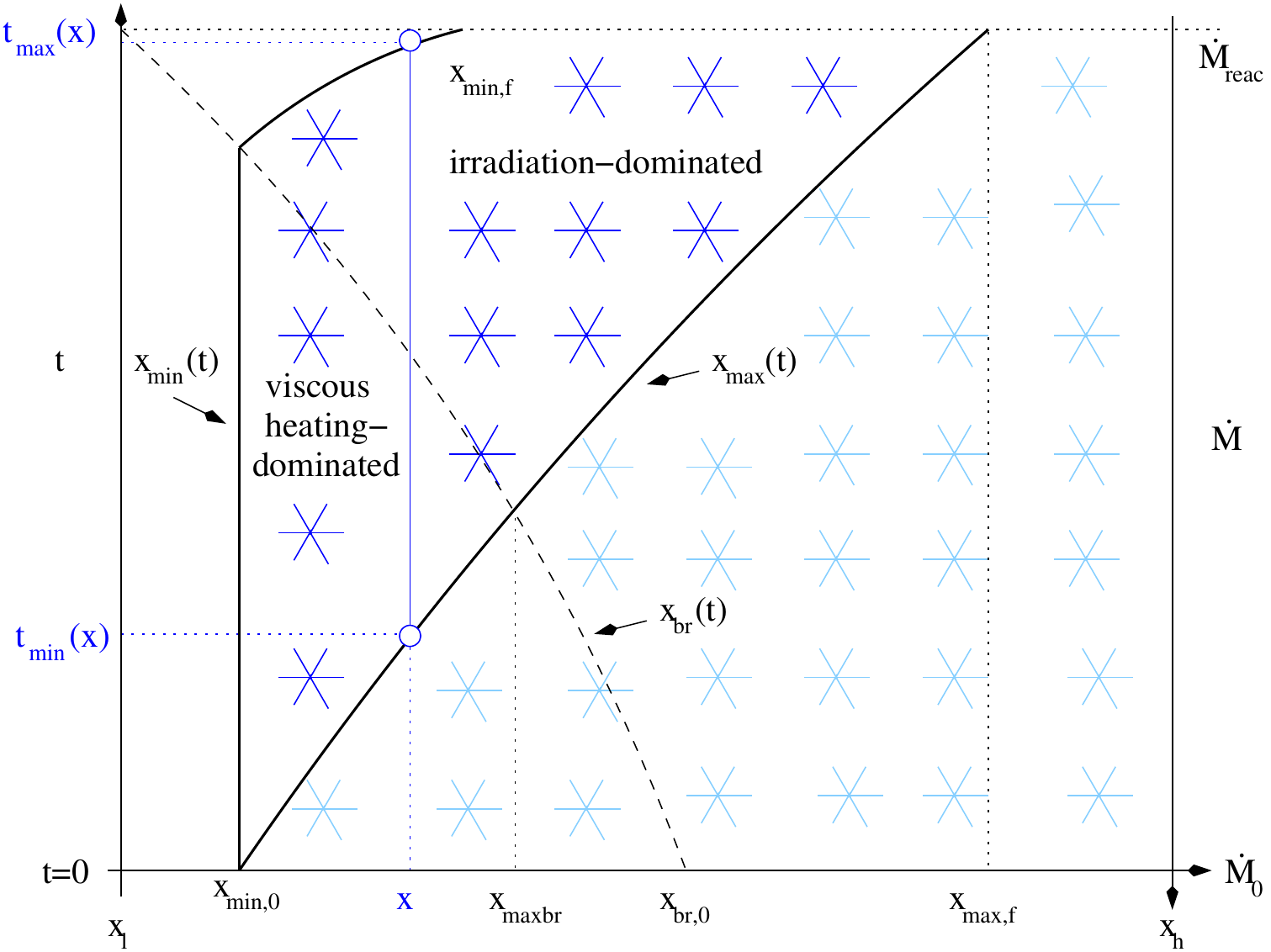}
}
\caption{A schematic diagram to help visualize the calculation and the different quantities involved. The abscissa is the D/H ratio ($x$) and the ordinate is the time $t$, or, correspondingly, the mass accretion rate $\dot{M}$. The snowflakes mark the domain where water is condensed and those colored in deep blue mark water ice inside heliocentric distance $R_{\rm max}$, which contributes to the PDF. Their area is bounded by the thick black curves ($x_{\rm min}$ and $x_{\rm max}$), while the dashed line marks the boundary between the viscous-heating and irradiation-dominated regimes (for the location corresponding having the D/H value of $x$ at time $t$). With the passage of time, the average D/H ratio of water increases. The blue line represents the time interval of integration on the right-hand-side of equation (\ref{PDF without time dependence}) for a given value of $x$.}
\label{diagramme calcul}
\end{figure}

  In this appendix, we detail the calculation of the PDF of the D/H ratio of meteoritic water. We will here denote D/H by $x$ for the sake of legibility of the equations.

 With the model set up in Section \ref{Modeling}, the mass of meteoritic water with D/H ratio between $x$ and $x+\mathrm{d}x$ is  
\begin{eqnarray}
\int_0^{t_{\rm reac}}
S_{\rm coll}(R(x,t),t)2\pi R(x,t)\frac{\partial R}{\partial x}_{|t}\mathrm{d}x\theta(R_{\rm max}-R(x,t))\mathrm{d}t\nonumber\\
\propto f(x)\mathrm{d}x
\label{f(x)}
\end{eqnarray}
where $f(x)$ is the normalized probability density function of bulk chondrites in terms of D/H ratio. $t=0$ is taken to correspond to the time where $R_{\rm cond}=R_{\rm max}$, i.e. the earliest possible time where water can condense and be accreted inside $R_{\rm max}$, corresponding to the mass accretion rate
\begin{eqnarray}
\dot{M}_0 & = & \left(\frac{128\pi^2\sigma_{\rm SB}k_B\alpha T_{\rm cond}^5}{3\kappa m \Omega(R_{\rm max})^3}\right)^{1/2}\nonumber\\
& = & 1.5\times 10^{-7}\:\mathrm{M_\odot.a^{-1}}\left(\frac{\alpha}{10^{-3}}\frac{0.5\:\rm m^2/kg}{\kappa}\right)^{1/2}\nonumber\\
& & \left(\frac{T_{\rm cond}}{170\:\rm K}\right)^{5/2}\left(\frac{R_{\rm max}}{10\:\rm AU}\right)^{9/4}
\end{eqnarray}
Using the assumed constancies of $\epsilon_{\rm H_2 O}$, $\Omega t_{\rm coag}(R)$, $\alpha$, Sc$_R$ and using equation (\ref{steady mass accretion rate}), we have:
\begin{eqnarray}
f(x) & \propto & \int_0^{t_{\rm reac}}\frac{\dot{M}(t)}{R_{\rm reac}(t)T(R(x,t),t)}\left(x_h-x\right)^{2/(3\textrm{\scriptsize Sc}_R)-1}\nonumber\\
& & \theta(R_{\rm max}-R(x,t))\theta(R(x,t)-R_{\rm cond}(t))\mathrm{d}t \nonumber\\
&\propto & \left(x_h-x\right)^{1/(3\textrm{\scriptsize Sc}_R)-1}\int_{t_{\rm min}(x)}^{t_{\rm max}(x)}\left(\frac{\dot{M}(t)}{\dot{M}_0}\right)^{7/9}\nonumber\\
& & \mathrm{min}\left(\left(\frac{\dot{M}_{\rm br}(x)}{\dot{M}(t)}\right)^{2/9},1\right)%\mathrm{min}\left(\left(\frac{\dot{M}(t)}{\dot{M}_{\rm reac}}\right)^{2/9},1\right)
\mathrm{d}t
\label{PDF without time dependence}
\end{eqnarray}
with $t_{\rm min}(x)$ and $t_{\rm max}(x)$ determined by the conditions $R<R_{\rm max}$ and $R>R_{\rm cond}$, respectively, $\dot{M}_{\rm br}(x)$ the mass accretion rate for which $x(R_{\rm br})=x$, where $R_{\rm br}$ is defined as the heliocentric distance of the transition between the viscous-heating- and the irradiation-dominated regimes and is given by (see equation (\ref{T steady})):
\begin{eqnarray}
R_{\rm br,AU} & = & \left(\frac{3}{128\pi^2}\frac{\kappa m}{\sigma_{\rm SB}k_B\alpha}\frac{\dot{M}^2\Omega_0^3}{(f_TT_0)^5}\right)^{1/2}\nonumber\\
& = & 3\:%\mathrm{AU}\:
\frac{\dot{M}_{-8}}{f_T^{5/2}}\left(\frac{\kappa}{0.5\:\mathrm{m^2/kg}}\frac{10^{-3}}{\alpha}\right)^{1/2}.
\label{Rbr}
\end{eqnarray}
with $\Omega_0\equiv \Omega(1\:\rm AU)$. % and $\dot{M}_{-8}\equiv \dot{M}/(10^{-8}\:\rm M_{\odot}\cdot a^{-1})$ %and corresponds to a temperature of
%\begin{equation}
%T_{\rm br}=\left(\frac{128\pi^2\sigma_{\rm SB}k_B\alpha (f_TT_0)^9}{3\kappa m \dot{M}^2\Omega_0^3}\right)^{1/4}
%= 150\:\mathrm{K} f_T^{9/4}\left(\frac{0.5\:\mathrm{m^2/kg}}{\kappa}\frac{\alpha}{10^{-3}}\right)^{1/4}\dot{M}_{-8}^{-1/2}
%\end{equation}
The %two
 ``min(...)'' factor in equation (\ref{PDF without time dependence}) is
 unity when %$R_{\rm reac}$ in the viscous-heating-dominated regime and 
$R(x,t)$ is in the irradiation-dominated regime.

  Before proceeding to the result, we coin a few additional notations and then proceed to the integration. Fig. \ref{diagramme calcul} may help the reader to visualize the situation in the D/H - time space.

  For a given mass accretion rate, the minimum and maximum value of D/H dictated by the conditions $R_{\rm cond}\leq R \leq R_{\rm max}$ are:
\begin{equation}
x_{\rm min}(\dot{M})=
\left\{\begin{array}{rrrr}
x_h-\left(x_h-x_l\right)\left(\frac{T_{\rm cond}}{T_{\rm reac}}\right)^{5\textrm{\scriptsize Sc}_R/3}\:\mathrm{if}\: T_{\rm br}\leq T_{\rm cond}\\
x_h-\left(x_h-x_l\right)\left(\left(\frac{T_{\rm cond}}{f_TT_0}\right)^3\left(\frac{3\kappa m \dot{M}^2\Omega_0^3}{128\pi^2\sigma_{\rm SB}k_B\alpha T_{\rm reac}^5}\right)^{1/3}\right)^{\textrm{\scriptsize Sc}_R}\\\:\mathrm{if}\:T_{\rm cond}<T_{\rm br}<T_{\rm reac}\\
x_h-\left(x_h-x_l\right)\left(\frac{T_{\rm cond}}{T_{\rm reac}}\right)^{3\textrm{\scriptsize Sc}_R}\:\mathrm{if}\: T_{\rm br}\geq T_{\rm reac}
\end{array}\right.
\label{xmin}
\end{equation}
and
\begin{equation}
x_{\rm max}(\dot{M})=x_h-\left(x_h-x_l\right)\left(\frac{R_{\rm reac}(\dot{M})}{R_{\rm max}}\right)^{3\textrm{\scriptsize Sc}_R/2}
\end{equation}

The D/H ratio at $R=R_{\rm br}$ (for $\dot{M}\geq\dot{M}_{\rm reac}$) is:
\begin{eqnarray}
x_{\rm br}(\dot{M})=x_h-\left(x_h-x_l\right)\left(\frac{128\pi^2\sigma_{\rm SB}k_B\alpha(f_TT_0)^9}{3\kappa m \dot{M}^2\Omega_0^3T_{\rm reac}^4}\right)^{5\textrm{\scriptsize Sc}_R/12}
\end{eqnarray}

To all these, a ``0'' subscript will be added if evaluated for $\dot{M}=\dot{M}_0$ and ``f'' for $\dot{M}=\dot{M}_{\rm reac}$. Hence,
\begin{equation}
x_{\rm min,0}=x_{\rm max, 0}=x_h-\left(x_h-x_l\right)\left(\frac{T_{\rm cond}}{T_{\rm reac}}\right)^{5\textrm{\scriptsize Sc}_R/3}
\label{xmin0}
\end{equation}
\begin{equation}
x_{\rm br,0}=x_h-(x_h-x_l)\left(\frac{T_{\rm irr}(R_{\rm max})^9}{T_{\rm reac}^4T_{\rm cond}^5}\right)^{5\textrm{\scriptsize Sc}_R/12}
\end{equation}
\begin{equation}
x_{\rm min,f}=x_h-\left(x_h-x_l\right)\left(\frac{T_{\rm cond}}{T_{\rm reac}}\right)^{3\textrm{\scriptsize Sc}_R}
\end{equation}
\begin{equation}
x_{\rm max,f}=x_h-(x_h-x_l)\left(\frac{T_{\rm irr}(R_{\rm max})}{T_{\rm reac}}\right)^{3\textrm{\scriptsize Sc}_R}
\label{xmaxf}
\end{equation}
where $T_{\rm irr}(R_{\rm max})\equiv f_TT_0R_{\rm max, AU}^{-1/2}$ is the irradiation temperature at heliocentric distance $R_{\rm max}$.

We will also need the D/H ratio when $R_{\rm br}=R_{\rm max}$, which is given by:
\begin{equation}
x_{\rm max br}=x_h-\left(x_h-x_l\right)\left(\frac{T_{\rm irr}(R_{\rm max})}{T_{\rm reac}}\right)^{5\textrm{\scriptsize Sc}_R/3}
\end{equation}
and the ratio $\dot{M}_{\rm reac}/\dot{M}_0$ between the final and the starting mass accretion rate:
\begin{equation}
\frac{\dot{M}_{\rm reac}}{\dot{M}_0}=\left(\frac{T_{\rm irr}(R_{\rm max})^9}{T_{\rm reac}^4T_{\rm cond}^5}\right)^{1/2}
\end{equation}

With all these notations, we can express the result of the integration in equation (\ref{PDF without time dependence}) as:
\begin{eqnarray}
f(x)=C(x_h-x)^{1/(3\textrm{\scriptsize Sc}_R)-1}\Bigg(2\left(\frac{x_h-x_{\rm br,0}}{x_h-x}\right)^{2/(15\textrm{\scriptsize Sc}_R)}\nonumber\\
-\frac{T_{\rm irr}(R_{\rm max})}{T_{\rm reac}^{13/18}T_{\rm cond}^{5/18}}\left(\frac{x_h-x_l}{x_h-x}\right)^{13/(30\textrm{\scriptsize Sc}_R)}-\left(\frac{\dot{M}_{\rm reac}}{\dot{M}_0}\right)^{1/9}S(x)\Bigg),
\end{eqnarray}
for $x_{\rm min,0}\leq x \leq x_{\rm maxbr}$, and
\begin{eqnarray}
f(x)=C(x_h-x)^{1/(3\textrm{\scriptsize Sc}_R)-1}\Bigg(\left(\frac{x_h-x}{x_h-x_{\rm min,0}}\right)^{1/(6\textrm{\scriptsize Sc}_R)}\nonumber\\
  -\left(\frac{\dot{M}_{\rm reac}}{\dot{M}_0}\right)^{1/9}S(x)\Bigg)
\end{eqnarray}
for $x_{\rm maxbr}\leq x \leq x_{\rm max,f}$. %Otherwise, $f(x)=0$.

  Here, $C$ is a normalization factor given by
\begin{eqnarray}
 & C  = \Bigg[\textrm{Sc}_R(x_h-x_l)^{1/(3\textrm{\scriptsize Sc}_R)}\bigg(
12\frac{T_{\rm irr}(R_{\rm max})^{1/2}T_{\rm cond}^{1/18}}{T_{\rm reac}^{5/9}}
-18\frac{T_{\rm irr}(R_{\rm max})^{5/6}}{T_{\rm reac}^{5/9}T_{\rm cond}^{5/18}}\nonumber\\
& +10\frac{T_{\rm irr}(R_{\rm max})}{T_{\rm cond}^{4/9}T_{\rm reac}^{5/9}}
+\frac{T_{\rm irr}(R_{\rm max})^{3/2}}{T_{\rm cond}^{5/18}T_{\rm reac}^{11/9}}
-\frac{T_{\rm irr}(R_{\rm max})^{1/2}T_{\rm cond}^{13/18}}{T_{\rm reac}^{11/9}}\bigg)\Bigg]^{-1}
\end{eqnarray}
and $S(x)$ is defined as
\begin{equation}
S(x)=
\left\{\begin{array}{rr}
\left(\frac{x_h-x}{x_h-x_{\rm min,f}}\right)^{1/(6\textrm{\scriptsize Sc}_R)}\:\mathrm{if}\:x<x_{\rm min,f}\\
1\:\:\:\:\:\:\:\:\mathrm{if}\:x\geq x_{\rm min,f}
\end{array}\right.
\end{equation}

\end{appendix}

\bibliographystyle{natbib}
\bibliography{bibliography}

\end{document}